\documentclass[a4paper]{aa}
\usepackage[dvips]{graphics}
\usepackage{graphicx}
\usepackage{xspace}
\usepackage{amssymb}
\usepackage{txfonts}

\newcommand{\ascc}   {\mbox{ASCC-2.5}\xspace}
\newcommand{\clucat} {\mbox{COCD}\xspace}

\newcommand{\dias}{\mbox{DLAM}\xspace}

\newcommand{\olin}[1]{\overline{#1}}
\newcommand{\veps}{\varepsilon}
\renewcommand{\ni}{\noindent}
\newcommand{\mc}[3]{\multicolumn{#1}{#2}{#3}}

\begin{document}

\titlerunning{Population of  Galactic open clusters}
\authorrunning{Piskunov et al.}

\title{Revisiting the population of Galactic open clusters}

\author{A.E.~Piskunov	\inst{1,2,3} \and
        N.V.~Kharchenko \inst{1,2,4} \and     
        S.~R\"{o}ser    \inst{2},
     	  E.~Schilbach	   \inst{2} \and
        R.-D.~Scholz	   \inst{1}
       }

\offprints{R.-D.~Scholz}

\institute{Astrophysikalisches Institut Potsdam, An der Sternwarte 16,\\
     	     D--14482 Potsdam, Germany\\
           email: apiskunov@aip.de, nkharchenko@aip.de, rdscholz@aip.de
	        \and
           Astronomisches Rechen-Institut, M\"onchhofstra{\ss}e 12-14,
           D--69120 Heidelberg, Germany\\
           email: apiskunov@ari.uni-heidelberg.de, elena@ari.uni-heidelberg.de,
                  nkhar@ari.uni-heidelberg.de,\\ roeser@ari.uni-heidelberg.de
           \and
           Institute of Astronomy of the Russian Acad. Sci.,
	        48 Pyatnitskaya Str., 109017 Moscow, Russia\\
	        email: piskunov@inasan.rssi.ru
           \and
           Main Astronomical Observatory, 27 Academica Zabolotnogo Str.,
           03680  Kiev, Ukraine\\
           email: nkhar@mao.kiev.ua
}

\date{Received 5 July 2005; accepted ...}
\abstract{\ni We present results of a study of the galactic open cluster
population based on the all-sky catalogue ASCC-2.5 (I/280A)  compiled from
Tycho-2, Hipparcos and other catalogues. The sample of optical clusters from
ASCC-2.5 is complete up to about 850 pc from the Sun.
The symmetry plane of the clusters' distribution is determined to be at $Z_0=-22\pm4$ pc,
and the scale height of open clusters is only $56\pm3$ pc. The total surface density
and volume density in the symmetry plane are $\Sigma=$ 114 kpc$^{-2}$ and
$D(Z_0)=1015$ kpc$^{-3}$, respectively. We find the total number of open clusters in
the Galactic disk to be of order of 10$^5$ at present.
Fluctuations in the spatial and velocity distributions are attributed to the existence
of four open cluster complexes (OCCs)  of different ages containing up to a few tens of
clusters. Members in an OCC show the same kinematic behaviour, and a narrow age spread.
We find, that the youngest cluster complex, OCC~1 ($\log t<7.9$), with
19 deg inclination to the Galactic plane, is apparently a  signature of Gould's
Belt. The most abundant OCC~2 complex has moderate age ($\log t\approx8.45$).
The clusters of the Perseus-Auriga group, having the same age as OCC~2, but 
different kinematics
are seen in breaks between Perseus-Auriga clouds. The oldest ($\log t\approx8.85$)
and sparsest group was identified due to a large motion in the Galactic anticentre
direction.
Formation rate and lifetime of open clusters are found to be
$0.23\pm0.03$ kpc$^{-2}$Myr$^{-1}$ and 322$\pm$31 Myr,
respectively. This implies a total number of cluster generations in
the history of the Galaxy between 30 to 40. We estimate that less than about 10\% of the 
total Galactic stellar disk population has ever passed an open cluster membership.
}


\maketitle

\section{Introduction}

Open clusters play a double role in astrophysical studies. Since
they consist of stars which  have been born and lived together, they 
provide natural laboratories to prove or rule out theories of the formation and evolution
of stars. Due to the fact that all members have about the same distance from the observers,
that they show common space motion, and have uniform chemical composition, each single
cluster represents excellent empirical reference sequences which are basic for
many contemporary scales and calibrations used in astrophysics.

On the other hand, studying the open cluster population provides important 
information that contributes to a better understanding of the structure and evolution of the
Milky Way.  In principal, basic parameters
like distance, motion, age, and metallicity can be determined for an open cluster
more accurately than for a single field star (roughly by a factor of $\sqrt{n}$ where $n$ is the number
of cluster members). Actually, they are better tracers of large scale structures
of the Galactic disk population than field stars.  
The most comprehensive  studies of the Galactic cluster population are about 20
years old (Lyng{\aa}~\cite{lyn82}, Janes et al.~\cite{janea88}). They were
based on the best data available at that time,  the
Lund Catalogue of Open Cluster Data (Lyng{\aa}~\cite{lyn87}, hereafter, the Lund Catalogue)
and its subset of  clusters with 3-colour photometry
(Janes \& Adler~\cite{janad82}). Although these studies present  an important
step in our understanding of the general properties of the cluster population,
they suffer from  incompleteness of the cluster samples and from inhomogeneity
of the clusters parameters. 

This can be illustrated by the following statistics. About
1200 clusters were known in the Lund catalogue by 1988. Only 400 of them had
accurate, but heterogeneous UBV photometry, and photometric distances, reddening
and age values. Although for almost all clusters apparent diameters were given
in the Lund catalogue (estimated by eye from sky charts or defined by the size
of detector's field of view), only about 100 clusters were
studied in a systematic way by use of stellar counts 
(Danilov \& Seleznev~\cite{dan}). Kinematics, being
the most traditional field in open cluster studies, was also hampered by heterogeneous
proper motions and radial velocities (RVs), and less than 100 clusters had
proper motions reduced to the fundamental reference system (van Schewick~\cite{schew}).
Space velocities were available for a few tens of clusters only.
Almost nothing was known about characteristic values of cluster statistics, 
how complete and representative the sample was, i.e. the basic parameters
needed for a systematic study of typical properties of the Galactic cluster population.
These difficulties were mainly due to the fact that data on various
clusters were strongly inhomogeneous since they were obtained by different
instruments, detectors and techniques, with  little effort to reduce them into 
one single system.

Since that time  considerable improvement has been achieved in the field of new and
homogeneous observations which could be used 
for the determination of basic cluster parameters and
for a systematic search for new clusters.
First of all, there are the
accurate all-sky surveys like Hipparcos/Tycho
in the optical and 2MASS in the infrared. The first efforts to look for open clusters in
these data led to the discovery of new objects both located in the Solar neighbourhood,
and at large distances from the Sun. Bica et al.~(\cite{biea03a})
compiled a list of 276 embedded infrared star clusters and stellar groups from 
the literature and were able to increase the number by 346 new compact and remote objects
associated with nebulae by making use of the 2MASS data (Dutra et al.~\cite{duea03},
Bica et al.~\cite{biea03b}). Also, nearby clusters were discovered in the
Hipparcos/Tycho catalogues. Platais et al.~(\cite{plea98}) found 15 nearby,
very loose and extended clusters and associations in the Hipparcos catalogue,
whereas Alessi et al.~(\cite{amd03}) discovered 11 new clusters within 0.8 kpc
with the Tycho data.

However, with respect to the astrophysical parameters for clusters, the situation had not much
improved during  recent years: the parameter set is neither complete nor
homogeneous. The on-line list of open cluster data by Dias et
al.~(\cite{dlam}, DLAM hereafter) which can be considered as a continuation
of the Lund Catalogue now contains by a factor  of 1.5 more clusters than its
predecessor, but the degree of completeness of this list is still unknown. Since
the cluster data in the DLAM list are taken from the literature (or from private
communications), the sets of the derived parameters differ from cluster to
cluster. Also, the parameters themselves are based on heterogeneous observations and
different methods of cluster definition and of parameter determination. Whenever using
these data for cluster population studies, one would meet problems caused
by uncertain cluster statistics and data heterogeneity.  Even one of the most
uniform lists of cluster parameters, the Loktin et al.~(\cite{lok01},\cite{lok04}) catalogue,
is based on non-uniform photometric data and uncertain membership. In other words,
for studies of the general properties of the Galactic cluster population, we cannot
simply benefit from a larger sample of open clusters as long as we cannot estimate its
homogeneity.

Ideal preconditions for cluster population studies are deep and uniform sky
surveys of sufficiently accurate astrometric and photometric data supplemented by
radial velocities, metallicities, and spectral classifications. Statistical properties of a
sample of clusters identified with these data could be well estimated since its
completeness should correlate with the completeness limit of the surveys used.
Applying uniform criteria and methods of membership and parameter determination,
one would get a homogeneous set of basic cluster parameters. In the near future
an important step can be undertaken in this direction when the 2MASS data will
be combined with the final version of the UCAC catalogue, and with the RAVE and
SEGUE surveys. In the further future, the observations of the GAIA satellite
will provide data at an even higher level of accuracy and homogeneity.

At the moment, more or less complete and accurate data are
available only for relatively bright stars and consequently for relatively nearby
clusters. Nevertheless, it is worth trying to use the present data for the re-determination
of cluster memberships and for the re-estimation of basic cluster parameters by applying 
uniform methods to homogeneous date sets.
This gives us a basis to increase
the reliability of the results on general properties of clusters, to estimate the 
possibilities and constraints of the data available, and therefore, to update our knowledge on 
the Galactic cluster population, now.

We started a long trip from the available observations to study the population of Galactic open
clusters with the compilation of a complete survey of stars and with the reduction
(if necessary) of the data to common reference systems. This effort resulted in
the All-Sky Compiled Catalogue of 2.5 million stars
(\ascc\footnote{\texttt{ftp://cdsarc.u-strasbg.fr/pub/cats/I/280A}}, Kharchenko~\cite{kha01})
with absolute proper motions in the Hipparcos system,
with $B$, $V$ magnitudes in the Johnson photometric system, and
supplemented with spectral types and radial velocities if available.
The \ascc was used to
identify known open clusters and compact associations from the Lund Catalogue,
the Dias et al.~(\cite{dlam}) on-line data collection,
and the Ruprecht et al.~(\cite{rupr}) list of associations. In the \ascc we
found 520 of about 1700 known clusters (Kharchenko et al.~\cite{starcat}, 
hereafter Paper~I) and discovered 130 new open clusters 
(Kharchenko et al.~\cite{newclu}, hereafter Paper~III). 
A pipeline was developed to
determine cluster membership based on kinematic and photometric criteria as well as
to obtain a uniform  set of cluster structural, kinematic and evolutionary
parameters (see paper I and Kharchenko et al.~\cite{clucat}, hereafter Paper~II).

In this paper we use the results of Papers II and III
to study unbiased properties of the cluster population
of the local Galactic disk. In \S~\ref{data_sec} we briefly describe the data set.
The completeness of the cluster sample and the spatial distribution of open clusters
in the Galactic plane and perpendicular to it is discussed in \S~\ref{backg_sec}.
The kinematics of the system of open clusters is considered in \S~\ref{kin_sec}.
Significant irregularity in the spatial and kinematic distributions of clusters
revealed complexes and groups of clusters which are discussed in \S~\ref{inhomo_sec}.
In \S~\ref{ltime_sec} we determine typical lifetime and formation rate of clusters
as it follows from the present statistics. Concluding remarks are given in
\S~\ref{concl_sec}.

\begin{figure*} 
\resizebox{\hsize}{!}{
\includegraphics[bb=95 210 580 625]{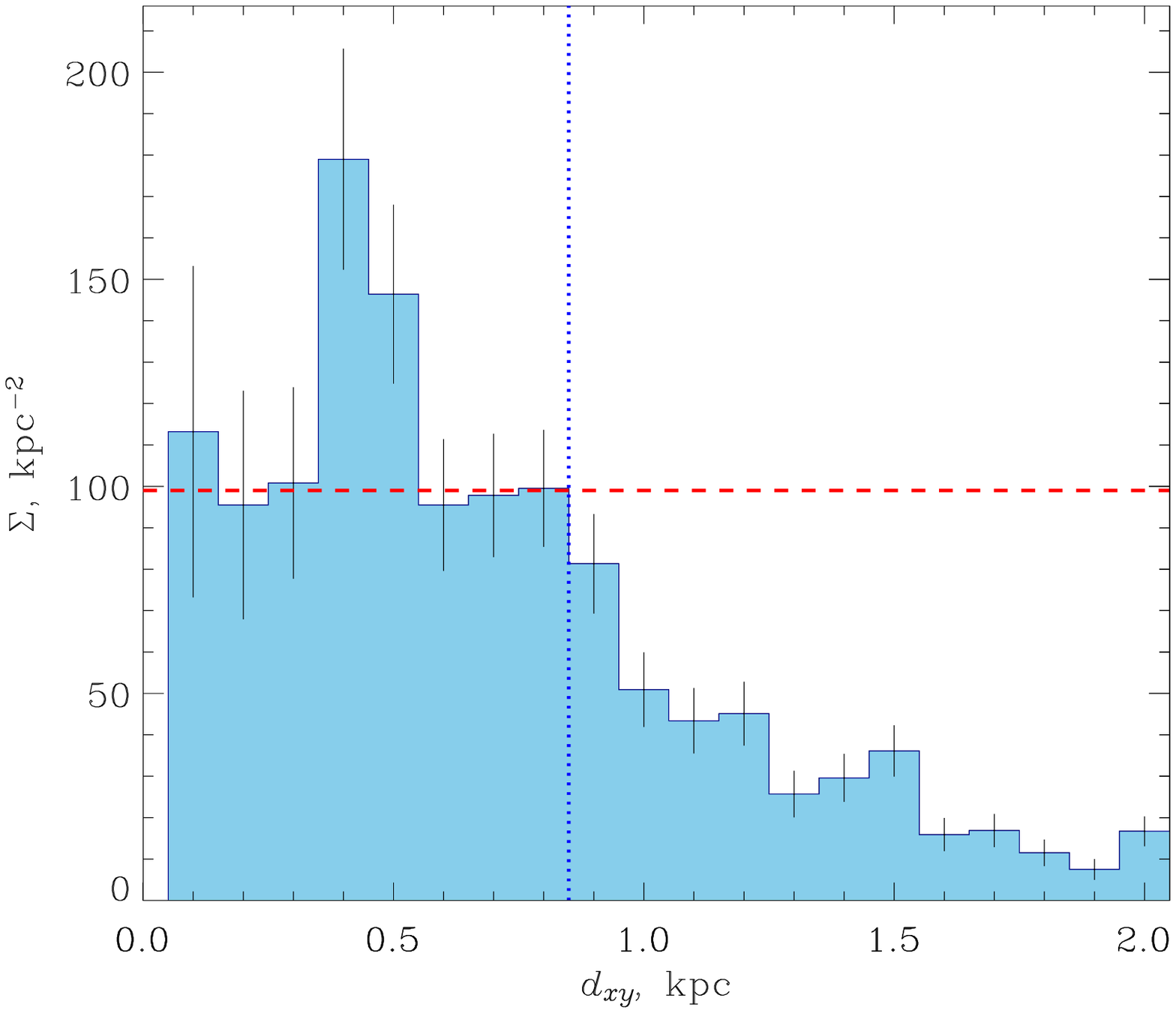}
\includegraphics[bb=65 210 580 630]{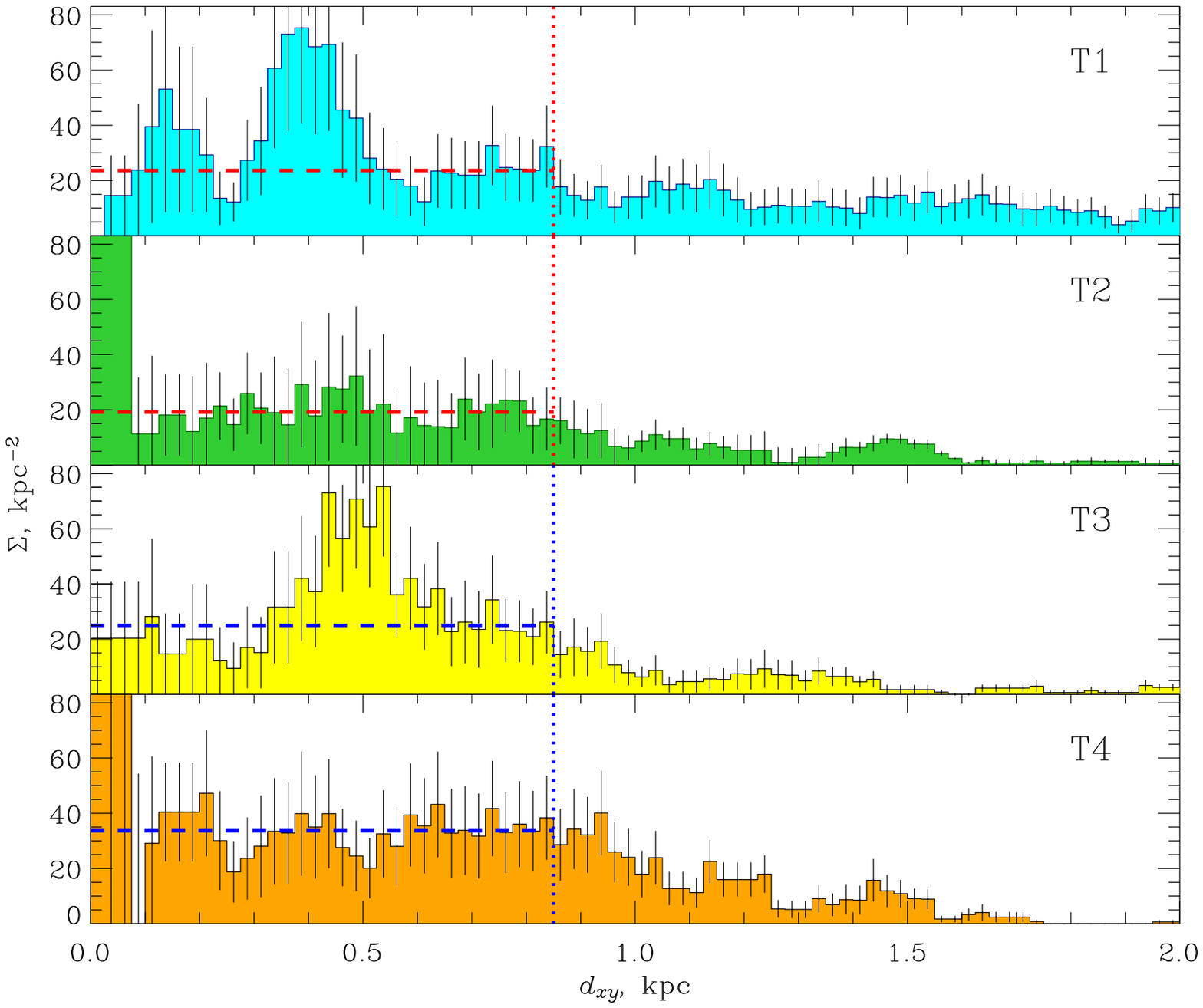}
}
\caption{Distribution of the surface density $\Sigma$ of open clusters versus
their distance $d_{xy}$ from the Sun projected onto the Galactic plane. The
dotted lines indicate the completeness limit, the dashed horizontal lines
correspond to the average density of ``field clusters'' (i.e., not contained in cluster
complexes, s. \S~\ref{inhomo_sec}). The bars are Poisson errors derived from cluster counts.
\textbf{Left panel:}  distribution of all clusters. \textbf{Right panels:}
distributions of clusters with different ages. The panel designations from T1 to
T4 are the identifiers of four subsamples of different ages as defined in the
text.
}\label{ddist_fig}
\end{figure*}

\section{Summary of the data used}\label{data_sec}

The present sample contains 520 already known clusters which we identified in
the ASCC-2.5 (Paper~I)
and 130 new open clusters which we newly detected in the \ascc data (Paper~III).
This sample of 650 open clusters is homogeneous in
many aspects: it is based on a practically uniform data set (i.e. the 
catalogue \ascc with proper motions and magnitudes reduced to uniform 
astrometric and photometric systems),
on a uniform approach of membership selection (kinematic and photometric
criteria, see Paper~I) and on a uniform technique of
the determination of cluster parameters applied in Papers II and III.
For each cluster in our sample, we determined distance and celestial position
(i.e. the location in 3D-space), reddening, size (including core and
corona radii), kinematics (the mean proper motions in the Hipparcos system), and
the cluster age computed from ages of individual members near the Main Sequence turn-off.
On the basis of a new cluster membership  determination, we derived radial
velocities and spatial velocity vectors for 359 clusters of our sample. For details
of the above parameter determination we refer the reader to Paper~II. 
The results are presented in two data sets:
The Catalogue of Open Cluster Data and its Extension~1. The catalogues are only available 
in electronic 
form at the CDS online archive\footnote{\texttt{ftp://cdsarc.u-strasbg.fr/pub/cats, 
http://vizier.u\mbox{-}\-strasbg.fr}}.
Hereafter, we refer to both  catalogues with the same identifier COCD. 

The two closest clusters, the Hyades and Collinder~285 (the UMa cluster), are
missing in our list. Due to their proximity to the Sun, they occupy large areas
on the sky and require a specific technique of membership determination.  In
order to complete our sample, we have taken the corresponding data for these
clusters from Loktin et al.~(\cite{lok04}) and the
WEBDA\footnote{\texttt{http://obswww.unige.ch/webda/}} database.  The mean 
proper motions and radial velocities for both clusters were computed with
data of the \ascc and 
CRVAD\footnote{\texttt{Catalogue of Radial Velocities with Astrometric Data 
can be retrieved from the CDS by ftp://cdsarc.u-strasbg.fr/pub/cats/III/239}} 
(Kharchenko et al.~\cite{kharv04}) catalogues. All our discussions in 
the following therefore refer to this sample of 652 clusters.

Although the current sample contains clusters covering a wide range of
ages  (from $\log t=6.6\ldots9.5$), it is strongly biased towards younger ages.
This is not surprising for an un-preselected cluster sample: due to dynamical evolution 
and evaporation of members, the typical life time of a cluster is of the order of a few hundred Myrs.
On the other hand, our cluster sample is magnitude limited. Since young clusters contain,
in general, more luminous stars, they dominate at larger distances and their proportion in the sample
is somewhat higher than that of older clusters.
Thus, in average our sample represents the moderately young disk.

\section{Spatial distribution in the disk}\label{backg_sec}

Throughout the paper we use the rectangular coordinate system $X,Y,Z$ with 
origin in the barycentre of the Solar system, and axes pointing to the Galactic centre
($X$), to the direction of Galactic rotation ($Y$), and to the North Galactic
pole ($Z$). The corresponding components of the space velocity are $U,V,W$, respectively.
Unless other specifications are given, we consider the $(X,Y)$ plane
to be the Galactic plane.

\begin{figure*}
\resizebox{\hsize}{!}
{\includegraphics[bb=60 210 535 660,clip=]{3764f02a.ps.gz}
\includegraphics[bb=60 210 535 660]{3764f02b.ps.gz}}
\caption{Open clusters and interstellar extinction. Distributions of the total 
extinction $A_V$ (left panel), and of the extinction coefficient $a_V$ (right panel)
in the wider Solar neighbourhood. In both panels the contours show $A_V,a_V$-levels
according to the grey-colour scales shown at the right upper corners of the panels.
The circles represent open clusters. Their sizes (and colour) correspond to $A_V,a_V$ of
the clusters. The blank areas in the periphery of the plots are due to a lack
of clusters in our sample at these locations.
}\label{av_fig}
\end{figure*}

\subsection{The completeness of the sample and the local density
enhancement}\label{compl_sec}
                  
In order to estimate the spatial completeness of the sample, we studied the
distribution of clusters in the $(X,Y)$ plane and computed their surface density
$\Sigma$ as a function of the projected distance $d_{xy}$ from the Sun.  The
corresponding histogram is shown in the left panel of Fig.~\ref{ddist_fig}. Due to the low number
of clusters at small distances from the Sun, the first bin represents  a  circle
with a radius of 150~pc. All other bins are concentric rings of 100~pc width each.
With the exception of a considerable excess at $d_{xy} = 0.35\ldots0.55$ kpc, the 
distribution is almost flat up to $d_{xy}=0.85$ kpc. At larger distances, the
surface density $\Sigma$ of clusters is steadily decreasing. Assuming a uniform 
density model for the distribution of clusters in the Solar neighbourhood, we
can interpret this behaviour as evidence for an increasing incompleteness of our
sample at projected distances larger than $0.85$ kpc. We adopt
$d^c_{xy}=0.85$ kpc, as the completeness limit of our sample.

On the other hand, the enhancement of the observed cluster density at 
$d_{xy}=0.35\ldots0.55$ kpc is too significant to be attributed to random fluctuations
in the spatial distribution of open clusters in the Solar neighbourhood.
Assuming a density of $100 \pm 10$ clusters per square kpc as typical of
the ``cluster field'' around the Sun, we would expect $56 \pm 8$ clusters at
$d_{xy}=0.35\ldots0.55$ kpc. Instead of this, 91 clusters are counted in this region.
Does this enhancement describe a general property of the local galactic 
structure or is this excess caused by a particular group (groups) of open clusters
with a common evolutionary history?

In order to understand the nature of this feature, we construct the surface
density distribution for samples of different ages. Altogether four cluster
generations are considered: the sample T1 including the 269 youngest clusters 
with ages $\log t \le 7.9$; T2 with 101 moderately young clusters of ages of
$\log t=(7.9,8.3]$; T3 with 132 intermediate age clusters of $\log t=(8.3,8.6]$;
T4 with 150 older clusters, $\log t>8.6$. The age
ranges are selected in such a way that the pattern of density distributions
reveals the highest contrast. For each sample we construct the histogram of the surface
density, smoothed with a five-point rectangular filter. The bin size is selected
to be 0.025 kpc. 

\begin{table*}
\centering
\caption[]{Distribution of the clusters along the $Z$-axis within a cylinder of
radius of $d^c_{xy}$.}
\label{zpar_tab}
\begin{tabular}{lcccccccc}
\hline
Parameter           & all        & all-``field''  & \mc{6}{c}{Cluster age samples}\\
 \cline{4-9}\rule{0mm}{3mm}                                         
                    &            &            &T1          &T1-``field''    &T2           &T3           &T3-``field''     &T4 \\
\hline
$Z_0$, pc           &$ -22\pm  4$&$ -20\pm  5$&$ -39\pm  8$&$ -29\pm  9$&$ -28\pm  9$ &$  -8\pm  8$ &$  -7\pm 11$ &$ -15\pm 10$\\
$N$                 &$	   259$  &$     211$  &$        72$&$        49$&$        45$ &$        76$ &$        51$ &$        66$\\
$D(Z_0)$, kpc$^{-3}$&$1015\pm 45$&$821\pm 40$&$ 263\pm 21$&$ 200\pm 20$&$ 209\pm 24$ &$ 376\pm 34$ &$ 232\pm 25$ &$ 236\pm 20$\\
$h_Z$, pc           &$  56\pm  3$&$  56\pm  3$&$  61\pm  6$&$  54\pm  6$&$  48\pm  6$ &$  44\pm  4$ &$  48\pm  6$ &$  61\pm  6$\\
$\sigma_Z$, pc      &$  74\pm  3$&$  75\pm  3$&$  72\pm  6$&$  68\pm  7$&$  63\pm  6$ &$  71\pm  5$ &$  78\pm  8$ &$  82\pm  7$\\
\hline
\end{tabular}
\end{table*}

The results are shown in the right panels of Fig.~\ref{ddist_fig}.
The horizontal dashed lines are average levels of
the ``field'' surface density $\Sigma_{0,i}$, determined  separately for each sample.
For the samples T2 and T4, they are simple averages of the densities within the
completeness area. The first bins in both groups, where by  chance one single 
nearby cluster produces a maximum of the distribution, are not taken into account
due to their low statistical
significance. For the samples T1 and T3 the ``field'' density is estimated from
different ranges of distance:  $\Sigma_{0,1}$  is computed from data in
$d_{xy}= 0.50\ldots0.85$ kpc, whereas for $\Sigma_{0,3}$ we consider a distance range of
$d_{xy}= 0.65\ldots0.85$ kpc. We discuss these groups in more detail in \S~\ref{inhomo_sec}.

\subsection{Open clusters and interstellar extinction}\label{av_sec}

Studies of the distribution of the obscuring matter in the Galactic disk 
have been carried out for many years (Sharov~\cite{shar63}, 
FitzGerald~\cite{fitz68}, Lucke~\cite{luck78}, Neckel \& Klare~\cite{neckl80},
Arenou et al.~\cite{aren92}). Usually, the methods make use of photometric data and spectral
classifications of field stars over the whole sky and provide both reddening and distance
evaluation. As a rule, these observations allow a mapping of interstellar extinction
within the local disk, typically up to 3 kpc.  Recently, Hipparcos
parallaxes and data on the equivalent widths of NaI lines in spectra of about 1000
local stars were used for a detailed mapping of the immediate ($d<0.3$ kpc)  Solar
neighbourhood (Lallement et al.~\cite{lal03}). As an upper limit of extinction
the maps constructed in Schlegel et al.~(\cite{sfd98}) can be taken, although they
are of limited use in the 1 kpc neighbourhood of the Sun.

Since open clusters reside in the Galactic disk, one must expect that they
are subject of sometimes strong and irregular interstellar extinction
which should be taken into account in their study. On the other hand,
open clusters having a compact structure, accurate distances, and with reddening
determined, as a rule, from a number of stars could be used as suitable probes
tracing the true distribution of reddening in the Galactic disk.
The purpose of this section is to reveal the important features of the distribution
of interstellar extinction from data on reddening of open clusters.

In Fig.~\ref{av_fig} we show the spatial distribution of open clusters together
with parameters describing the extinction in $V$: the total extinction $A_V$ 
and the extinction coefficient  $a_V= A_V/d$.
Since $A_V$ characterises the total amount  of absorbing
matter along the line of sight, the total extinction is increasing with
distance. One observes this increase towards to the periphery of
Fig.~\ref{av_fig} (left panel), and notes its different behaviour in different
directions. Generally, the $A_V$ distribution can be used for examining the
visibility conditions in a given direction and their apparent trends. The
extinction coefficient is related to dust and can be used for mapping dust clouds 
(Fig.~\ref{av_fig}, right panel).

The total extinction was computed in quadratic areas of 75$\times$75~pc$^2$
and then smoothed via a 5$\times$5 rectangular window i.e., the pattern
displayed is averaged over a scale of about 0.4 kpc. In contrast to this, for better
representation of cloud structures, we smoothed the maximum values of $a_V$
(instead of their averages like for $A_V$) which were determined in each square
of 75$\times$75~pc$^2$.

In Fig.~\ref{av_fig} we show, also, all clusters of our sample with 
individual values of $A_V$ and $a_V$ indicated by symbol size (and colour). The
same size (colour) scale is used in both panels: the smallest (white) circles
mark open clusters with $A_V < 0.5$ mag (left panel) or $a_V < 0.5$ mag/kpc
(right panel); the small (yellow), medium (dark yellow), large (red) and the 
largest (brown, dotted) circles indicate clusters with $A_V$/$a_V$ ranging into
$(0.5,1.5],\, (1.5,2.5],\, (2.5,3.5]$, and $>3.5$ mag or mag/kpc, respectively.

Some general features of the extinction pattern can be observed in the left panel
of Fig.~\ref{av_fig}. At the location of the Sun we find a 
cavity with a small extinction of the order of $0\fm1\ldots0\fm2$, which is transformed
into a long transparent ($A_V\approx0\fm2\ldots0\fm3$) corridor located in the directions
$l\approx80^\circ$ and 220$^\circ$ of galactic longitude. Another corridor of about the
same transparency ($A_V\approx 0\fm2$) could be seen if we consider only clusters 
with $Z<-20$ pc. It is directed towards 
$l\approx120^\circ$ and $l\approx300^\circ$ from the Sun.

Another feature is the complete absence of clusters (and also extinction data) in
some more peripheral areas. Most interesting are the areas at $(X,Y)\approx(-1.2,1)$ kpc
and at $(X,Y)\approx(1.2,1)$ kpc which manifests a screening effect of massive 
nearby clouds. Some of these clouds are clearly seen in the right panel, and they
are identified with the Auriga-Taurus clouds at $X<-0.1$ kpc, and  the Ophiuchus
clouds at $X\ge0$ kpc. However, the empty area at $(X,Y)\approx(1.2,1)$ kpc is not
caused by the Ophiuchus clouds. Instead, the molecular clouds located in the
Aquila Rift region play a more important role here. The clouds are extended farther
in the direction of  $l\approx50^\circ$ 
(Dame et al.~\cite{damea01}), and are not seen in our map because the known
clusters in this direction are too faint in the optical to be identified in the \ascc.

The impact of the clouds on the visible distribution of the clusters is not limited to
the Solar vicinity, only. Their shadow can be seen at larger distances and they influence
locally the visibility of such structures as the Sagittarius and Perseus arms at some kpc away
from the Sun (see \S~\ref{spir_sec}). Although the spatial density of open clusters is
lower than the density of individual stars, the extinction pattern obtained is in a good
agreement with most of the previous publications based on data of field stars.

\begin{figure}
\resizebox{\hsize}{!}
{\includegraphics[bb=73 66 563 516,origin=c,angle=270,clip=]{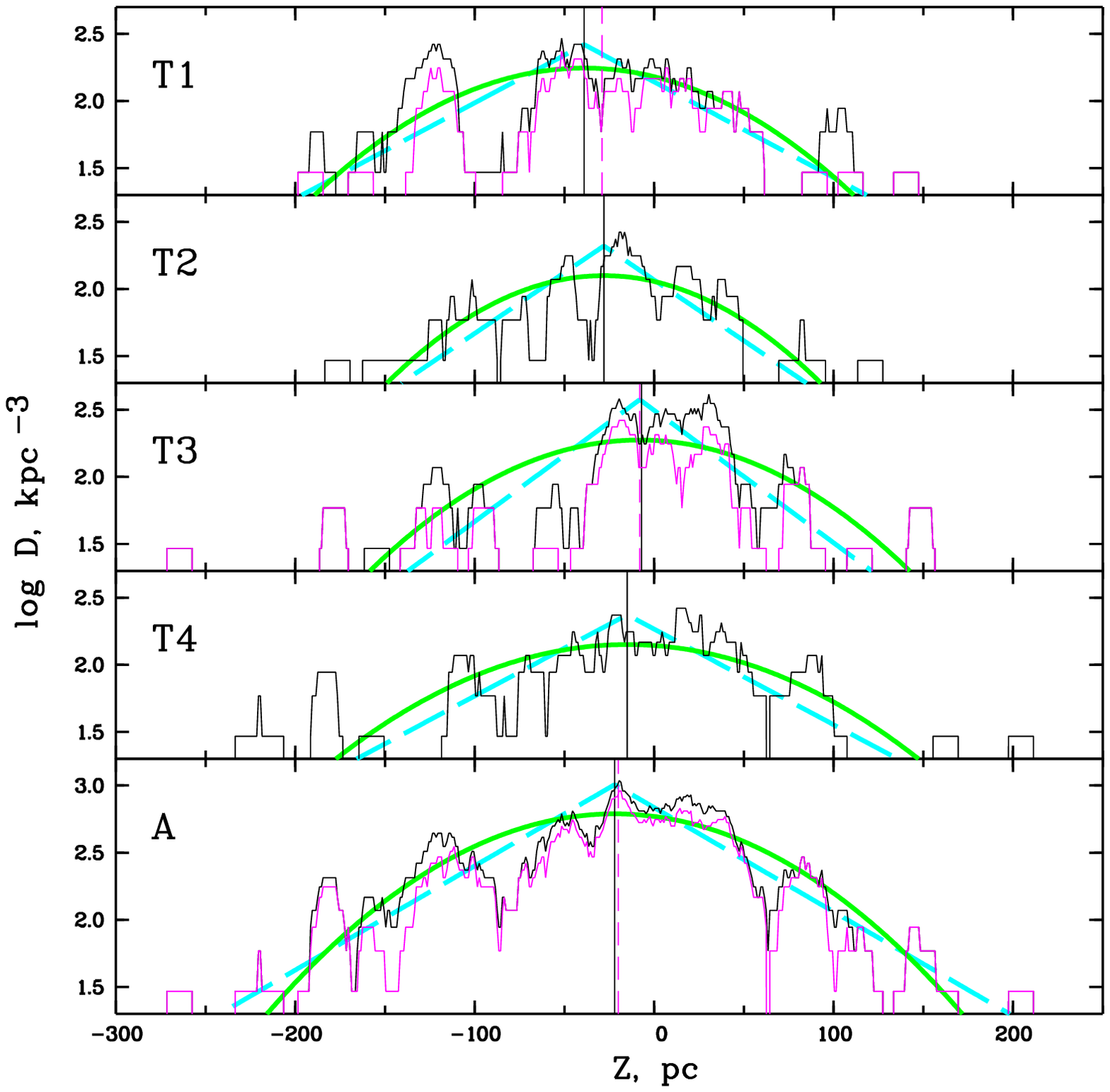}
}
\caption{Distribution of clusters perpendicular to the Galactic plane.
The distributions are constructed for clusters within the completeness limit of 850~pc. 
Panel~A: all clusters, panels T1 - T4: the 4 age samples (from T1 to T4) defined
in \S~\ref{compl_sec}. 
The thin black curves indicate smoothed spatial densities. The thin
(magenta) curves in panels (T1),(T3) and (A) mark
``field'' subsamples of the corresponding ages. The vertical solid lines give positions ($Z_0$)
 of symmetry planes, and the dashed vertical lines show $Z_0$ of ``field'' subsamples.
The thick solid curves correspond to Gaussian distributions computed with parameters
listed in Table~\ref{zpar_tab}. Thick long-dashed lines give the barometric distributions
fitted to observations.
}\label{zdist_fig}
\end{figure}

\subsection{Distribution of clusters perpendicular to the Galactic plane}\label{z_sec}

Distribution along the $Z$-axis is usually described in terms of the barometric
formula given in eq.~(\ref{denz_eqn}) where $D(Z_0)$ is the density at the position
of the symmetry plane $Z_0$ and $h_Z$ is the scale height:
\begin{equation}
D(Z)=D(Z_0)\exp\left\{{-\frac{|Z-Z_0|}{h_Z}}\right\}. \label{denz_eqn}
\end{equation}
The distribution parameters $Z_0$, $h_Z$ and $D(Z_0)$ were computed with data on 
259 clusters located within the completeness limit determined in 
\S~\ref{compl_sec}. Additionally, we considered the distributions of clusters of
the age groups defined in \S~\ref{compl_sec}. Further, we separated ``field'' subsamples
i.e., excluded probable kinematic members of the cluster complexes discussed in
\S~\ref{inhomo_sec}. The corresponding distributions along the $Z$-axis were 
constructed with a step of 1 pc and then smoothed by use of a rectangular 15-point filter.
The results are shown in Fig~\ref{zdist_fig}.

Due to poor statistics (the number of clusters in the different groups varies
from 45 to 259) and large density fluctuations, a simultaneous solution of
eq.~(\ref{denz_eqn}) did not yield a stable and evident result. Therefore,
we computed the distribution parameters in two steps. At first, the position of
the symmetry plane was assumed to be a simple average in each group. The medians
and modes were also considered but the results did not differ significantly.
For each sample, the standard deviation $\sigma_Z$ of $|Z-Z_0|$ was determined
as a statistical measure of cluster scattering. In a second step, the
parameters $D(Z_0)$ and $h_Z$ were derived by use of the integrated form of
eq.~(\ref{denz_eqn}):
\begin{equation}
\Sigma(Z)=2\,D(Z_0)\,h_Z\,\left(1-\exp\left\{{-\frac{|Z-Z_0|}{h_Z}}\right\}\right),\label{idenz_eqn}
\end{equation}
where $\Sigma(Z)$ is the surface density of clusters in a layer with 
width $2\,|Z-Z_0|$. The parameters were found from a non-linear fit of eq.~(\ref{idenz_eqn})
to the observed distributions. The results with corresponding $rms$ errors are 
given in Table~\ref{zpar_tab}.

\begin{figure*}
\resizebox{\hsize}{!}
{\includegraphics[bb=45 35 570 750,angle=270,clip=]{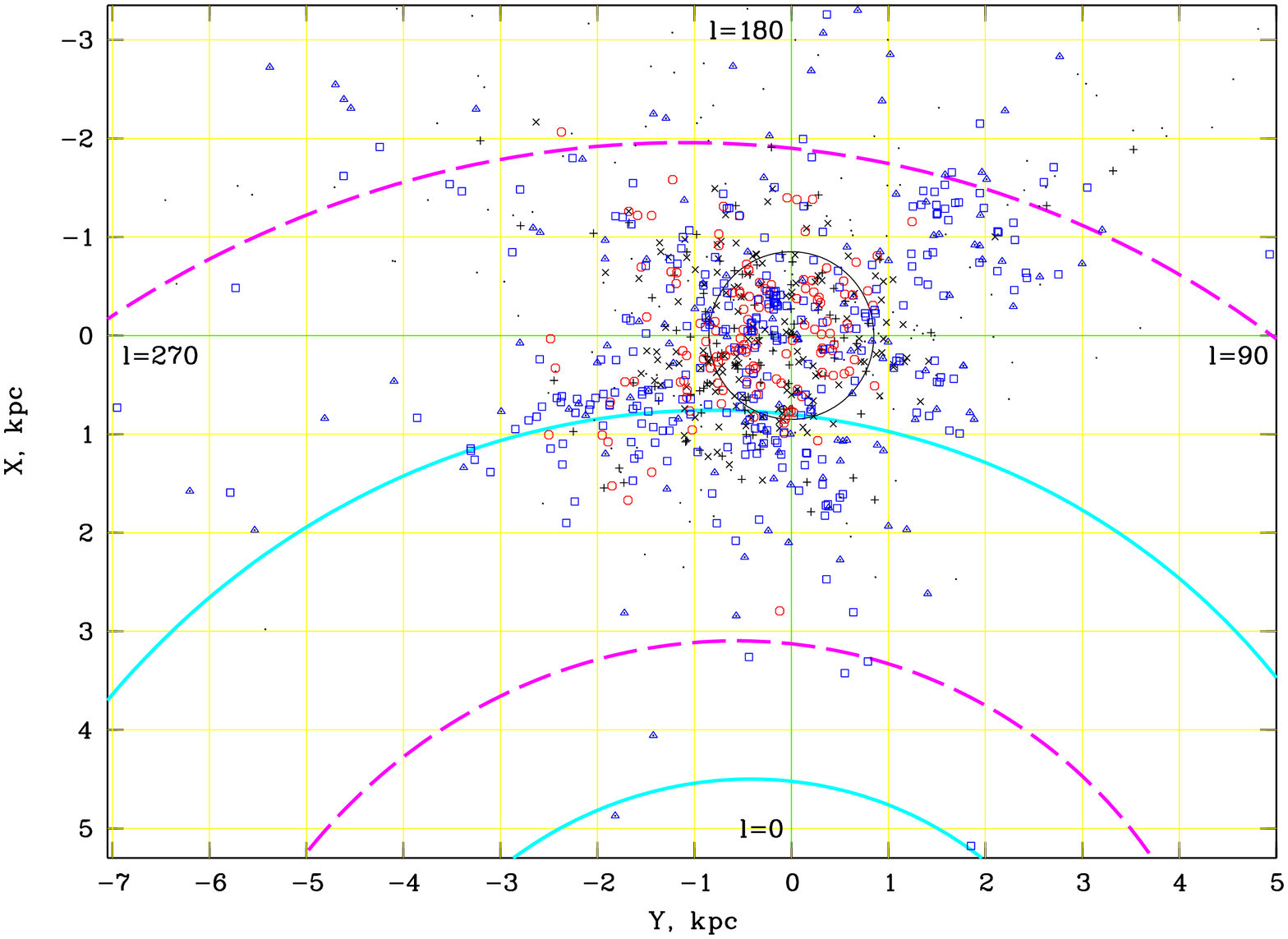}}
\caption{Distribution of open clusters in the $XY$-plane centered at the location of the
Sun. Blue squares mark T1 clusters,  black plusses -- T2,  red
circles -- T3,  and black crosses -- T4 clusters. Dots indicate the positions of 
clusters with known distances and ages from the \dias catalogue (v2.0), which are not included in
our sample. The youngest ($\log t <7.9$) of these clusters are marked by 
triangles. Long dashed and solid thick curves are the grand design spiral arms with
a pitch-angle of $-6^\circ$ fitted to young clusters in the Perseus and Carina-Sagittarius 
regions. The large circle around the Sun indicates the completeness area of our sample.}
\label{xy_fig}
\end{figure*}

According to  Fig.~\ref{zdist_fig} and Table~\ref{zpar_tab}, open clusters
form an extremely flat system in the Solar neighbourhood. The distribution
demonstrates non-regular behaviour with many peaks and gaps. We attribute this to
a possible impact of interstellar clouds on the apparent cluster distribution
(see also \S~\ref{av_sec}). Although varying from one sample to another, 
the symmetry plane of the cluster system is located below the $(X,Y)$ plane.
This finding indicates the position of the Sun above the physical plane of 
the system of open clusters. 
The derived displacement  $Z_0=22\pm4$ pc is in good agreement with
recent results obtained from studies of other young Population~I objects: according to 
Reed~(\cite{reed}) the position of the symmetry plane of OB stars within 1200 pc from the 
Sun is located at $Z_0=19.5\pm2.2$ pc, whereas the symmetry plane of the dust layer  
is at $Z_0=22.8\pm3.3$ pc (Joshi~\cite{joshi}).
The scale heights  vary within 50-60~pc without
an evident trend with respect to the ages of clusters (the youngest (T1) and the
oldest (T4) groups have the largest $h_Z$ value). Although the
standard deviations $\sigma_Z$ are always larger than the scale heights,
the corresponding Gaussian and barometric distributions do not
differ significantly (see Fig.~\ref{zdist_fig}).  

We found the total volume and surface densities of clusters in the plane of symmetry to be
$D(Z_0)=1015$ kpc$^{-3}$ and $\Sigma=$ 114
kpc$^{-2}$, respectively. The latter number considerably (by a factor of about 5) exceeds the
surface density of open clusters which is inferred as $\Sigma =$24.9 kpc$^{-2}$
from Janes et al.~(\cite{janea88}, Table~VIII). 
This discrepancy can be explained by mainly two reasons: a) a significant number
of clusters within 0.85~kpc in our sample is missing in the sample of 
Janes et al.~(\cite{janea88}), and b) the completeness of 
their cluster sample is overestimated by Janes et al.~(\cite{janea88}). 
A combination of these effects should increase their result for the surface density
by a factor of 3-6.

Provided that the cluster density distribution in the Galactic disk can be
described by a radial density profile with a scale length of the disk $h_l$,
we estimate the total number of clusters $N_{tot}$  in the Galaxy as
\begin{equation}
N_{tot} = 2\,\pi\,\Sigma\,\int\limits_0^{R_G^{lim}} \exp \left( - \frac{R_G - R_{G\odot}}{h_l} \right)\,R_G\,dR_G.
\label{densp_eqn}
\end{equation}
Here $R_G$ is the distance from the centre of the Galaxy, $R_G^{lim}$ is the
radius of the  Galactic disk, and $h_l$ is the scale length,
adopted from Bahcall \& Soneira~(\cite{bahson}) to be $h_l=3.5$ kpc. With
$R_G^{lim} =$ 15 kpc, we derive the total number of open clusters in the Galaxy as 
$N_{tot}$ = 93\,000. 
Within a reasonable range of the parameters $R_G^{lim}=13-20$ kpc, and
$h_l>3$ kpc the above estimate is valid within 20\%. For smaller values of $h_l$,
the expected number of open clusters would come out to be considerably larger.
 
\subsection{Distribution of open clusters in the Galactic plane}\label{spir_sec}

The distribution of open clusters in the Galactic plane is presented in
Fig.~\ref{xy_fig}. There we show also some 300 clusters from the catalog \dias (version 2.0)
which have distances and ages, but are missing in our sample since
they are mostly too faint to be identified in the \ascc data. Only about 30 of them
are located within the completeness area of our sample. It turned out that all these
30 clusters did not fulfill our criteria for being real open clusters and were thus
excluded from the sample. The retained clusters shown in Fig.~\ref{xy_fig}
append the current distribution at larger distances.

Until now the pattern of the spiral structure of the Galaxy is not clearly established.
In order to fit the observations, different models have been proposed which suggest
two, three or four spiral arms,  with pitch angles ranging from $-5^\circ$ to $-21^\circ$
(see Val\'ee~\cite{val95} for reference). Also, many attempts were undertaken to discuss 
the apparent distribution of open clusters in the Galactic plane with their 
relation to a local spiral pattern
(see e.g. Becker~\cite{beck63}, Janes et al.~\cite{janea88}).
Since the clusters of our sample have homogeneous distances and ages, we can try to fit
their distribution by a spiral pattern, but we must take into account that the
real distribution of open clusters in the optical is hidden by lumpy extinction, and
we see distant clusters only by  chance through transparent windows
(cf. Figs.~\ref{av_fig} and~\ref{xy_fig}).

In Fig.~\ref{xy_fig} we show the grand design pattern of logarithmic spirals with
a pitch angle $p=-6^\circ$ fitted to the Perseus and Carina-Sagittarius arms. The
choice of the pitch-angle is dictated by our intention to provide a reasonable fit
of a simple pattern (i.e., a two-arm spiral) to the observed distribution of young clusters
at different galactocentric distances. We have succeeded in getting a reasonable
agreement along the whole Perseus arm (from $l\approx90^\circ$ to about $270^\circ$),
the Carina-Sagittarius arm (from $l\approx300^\circ$ to about $360^\circ$), and
its extension in Vulpecula at about $(X,Y)=(1,1.5)$ kpc. Taking into account
relative errors of distances of about 15\%, it is not excluded that
a few distant clusters at about X=3 kpc and X=5 kpc trace the inner convolutions
of the Perseus and Carina-Sagittarius spirals.

\begin{table}
\centering
\caption[]{Kinematic and orbital parameters of the sub-system of open clusters
in the Galaxy}
\label{kinem_tab}
\begin{tabular}{lcc}
\hline
Parameter                   &Value&$rms$ error\\
\hline                                                     
$U_\odot$, km/s      	       &$+ 9.44    $&$1.14   $\\
$V_\odot$, km/s      	       &$+11.90    $&$0.72   $\\
$W_\odot$, km/s      	       &$ +7.20    $&$0.42   $\\
$\sigma_U$, km/s               &$ 13.86    $&$0.81   $\\
$\sigma_V$, km/s               &$  8.75    $&$0.51   $\\
$\sigma_W$, km/s               &$  5.05    $&$0.30   $\\
$\sigma_U :\sigma_V : \sigma_W$&$1 : 0.63 : 0.36$&$  $\\
$\overline{R_a}$, kpc          &$ 8.631   $&$0.034   $\\
$\sigma_{R_a}$, kpc            &$ 0.427   $&$0.024   $\\
$\overline{R_p}$, kpc          &$ 6.706   $&$0.056   $\\
$\sigma_{R_p}$, kpc            &$ 0.682   $&$0.040   $\\
$\overline{Z_{max}}$, kpc      &$ 0.260   $&$0.016   $\\
$\sigma_{Z_{max}}$, kpc        &$ 0.189   $&$0.011   $\\
$\overline e$                  &$ 0.127   $&$0.003   $\\
$\sigma_e$                     &$ 0.037   $&$0.002   $\\
$A$, km/s/kpc                  &$+14.5$   &$0.8$   \\
$B$, km/s/kpc                  &$-13.0$   &$1.1$   \\
\hline
\end{tabular}
\end{table}

No open clusters are visible at the expected position of the Carina-Sagittarius
arm at $l=20-55^\circ$. We explain this by massive molecular clouds located in
this direction (see the discussion before). According to Bica et
al.~(\cite{biea03b}) the gap is well filled by 45 embedded clusters detected in
the 2MASS atlas. However, they cannot be used as tracers of the spiral arm since
reliable distances are still unknown for them.

\section{Kinematics of open clusters}\label{kin_sec}

The first efforts to transform known proper motions of open clusters to a common 
reference system was undertaken by van Schewick~(\cite{schew}) who
derived absolute proper motions of about 70 open clusters in the FK4 system.
Lyng{\aa}~(\cite{lyn82}) provided radial velocities for 108 open clusters, but
only for a few tens  of them all three vectors of the space velocity could be 
obtained at that time.
Due to the Hipparcos observations and to the development of coravel-type spectrovelocimeters,
this situation has improved considerably. Currently, all 652 clusters of our
sample have accurate proper motions in the Hipparcos system and uniform heliocentric 
distances. For about 55\% of them, the mean radial velocities ($RV$) are also 
available (Paper~II).
Since the kinematic data of clusters are generally more accurate than these of single
field stars, we make use of this advantage  to learn about the kinematics of the 
Galactic disk, in more detail. On the other hand, we should keep in mind that
the subsample of open clusters with complete kinematic data is biased towards younger 
clusters: $RV$s are available for about 75\% of clusters younger than $\log t=8.3$,
whereas for older clusters, radial velocities are known for only 31\% of them.
This has to be taken into account if one compares the kinematic parameters derived
from proper motions (or tangential velocities) of 652 clusters or from the
subsample of clusters with proper motions and radial velocities.

In the following, we distinguish three types of space velocities. Firstly, these are
observed velocities which are computed directly from \clucat proper motions
and radial velocities. For each cluster, this space velocity contains  systematic components due to Galactic     
differential rotation and due to Solar motion with respect to the centroid of the
open cluster system, as well as a peculiar motion of a cluster with respect to this centroid.
The second type are velocities corrected for Galactic differential rotation, and, finally,
the third type are velocities corrected both for Galactic differential rotation and for
Solar motion (i.e., the peculiar velocities of clusters with respect to the centroid
of the open cluster system).

As a first step, we determine kinematic parameters describing the 
basic motions of the system of open clusters in the Galactic disk: the motion 
with respect to the Sun, and the differential rotation around the Galactic centre.
The Solar motion components $U_\odot, V_\odot, W_\odot$ were derived with the
complete kinematic data of 148 clusters located within $d^c_{xy} \le$ 0.85 pc as:
\begin{equation}
U_\odot = -\olin{U},\;\; V_\odot = -\olin{V},\;\; W_\odot = - \olin{W}, \label{uvwsun_eqn}
\end{equation}
\ni with $U, V, W$ computed from proper motions and radial velocities with
\begin{eqnarray}
U &=&k\,d\,(\mu_l\,\sin l+\mu_b\,\cos l\,\sin b)-RV\,\cos l\,\cos b, \nonumber\\
V &=&k\,d\,(-\mu_l\,\cos l+\mu_b\,\sin l\,\sin b)-RV\,\sin l\,\cos b, \label{uvw_eqn}\\
W &=& -k\,d\,\mu_b\,\cos b-RV\,\sin b, \nonumber
\end{eqnarray}
where $k$ is the scale factor converting angular motions into linear velocities.
The Oort's constants $A$ and $B$ of the Galactic rotation were determined from
the proper motions $\mu_l, \mu_b$ of 581 clusters with distances less than
$d_{xy} \le$ 2500 pc by a LSQ-solution of the equations:

\begin{eqnarray}  
k\,d\,\mu_l&=&(U_\odot\,\sin l-V_\odot\,\cos l)+(A\,\cos 2l+B)\,\cos b, \nonumber\\
k\,d\,\mu_b&=&(U_\odot\,\sin b\,\cos l + V_\odot\,\sin b\,\sin l-W_\odot\,\cos b)- \nonumber\\ 
            &&\frac{A}{2}\,\sin 2b\,\sin 2l.   \label{galrot_eqn}           
\end{eqnarray}
A rigorous solution of eq.~(\ref{uvw_eqn}) assumes  velocity components
corrected for Galactic rotation, whereas the  velocities in
eq.~(\ref{galrot_eqn}) should be free from the Solar motion. In order to fit
these introduced corrections separately, we solved the equation system 
(\ref{uvwsun_eqn})-(\ref{galrot_eqn}) by iteration. Only a few iterations were
necessary before convergence was achieved.

In the next step, the space velocities of 148 clusters within $d_{xy} \le$ 0.85 kpc 
corrected for Solar motion and Galactic rotation 
were used for computing the parameters ($\sigma_U, \sigma_V, \sigma_W$) of the
velocity ellipsoid. Assuming the form of the Galactic potential proposed by
Saio \& Yoshii~(\cite{saio}), we also derived  elements of their Galactic box orbits. 
This includes the apocentre and pericentre distances $R_a$ and $R_p$, the eccentricity $e$
and the maximum vertical distance $Z_{max}$ which a cluster can reach in its 
orbital motion. The results are given in Table~\ref{kinem_tab}.

\begin{figure*}
 \resizebox{\hsize}{!}
{\includegraphics[bb=130 52 540 719,angle=270,clip=]{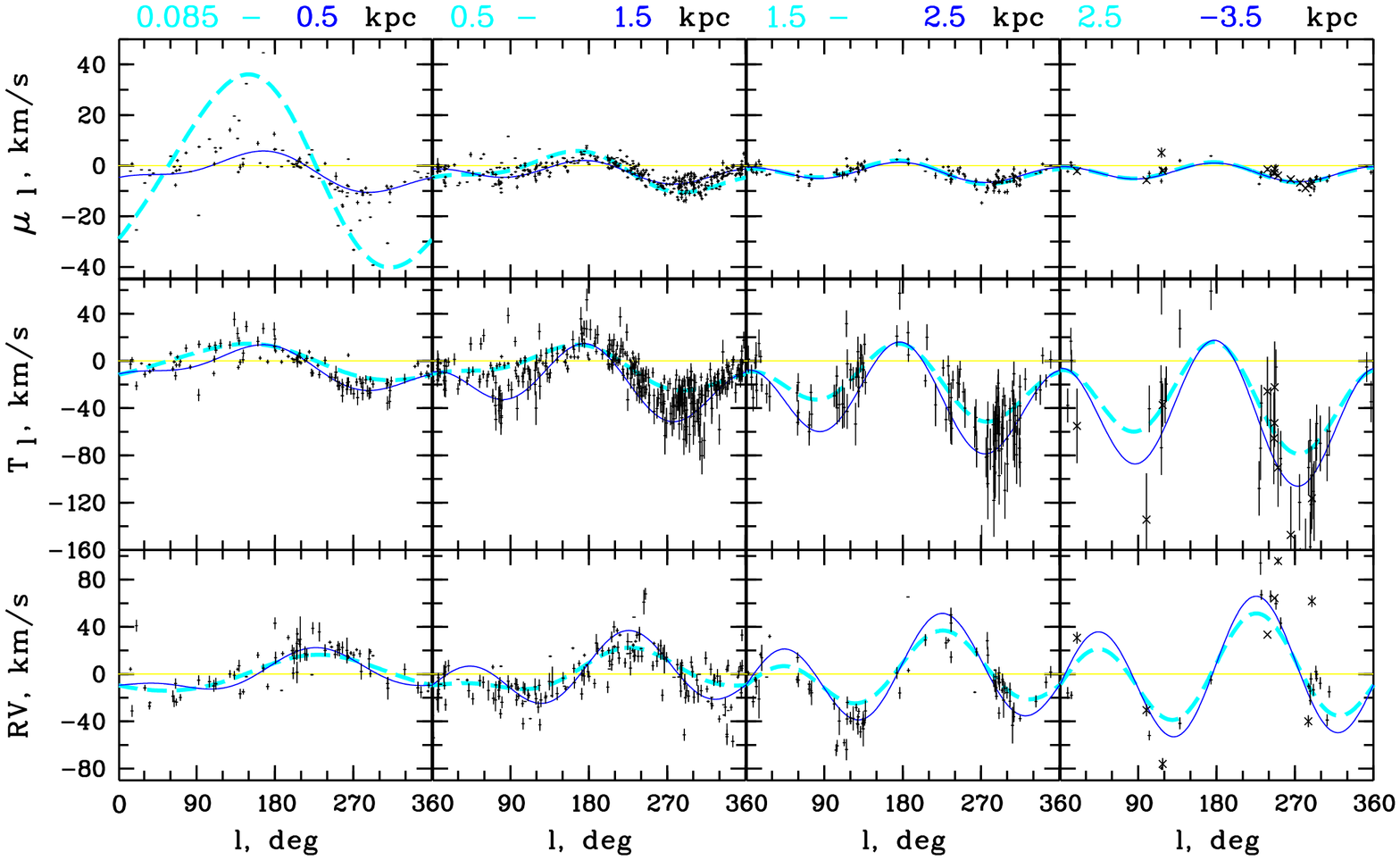}}
\caption{Observed kinematic data of open clusters versus Galactic longitude for
four ranges of distances (as indicated at the top). The upper panels are for  proper motions,
the middle - for tangential velocities, and the bottom - for radial velocities.
The parameters of individual clusters are marked by thin horizontal strokes together
with the vertical error bars (in the $RV$ panel the error bars are given only for clusters
for which the $rms$-errors could be estimated). The crosses in the rightmost panels
indicate clusters with $d >$ 3.5 kpc. For a given distance range, the curves show
systematic velocity components due to Solar motion and Galactic rotation
computed with the parameters $U_{\odot}, V_{\odot}, W_{\odot}$, A and B in 
Table~\ref{kinem_tab}. The thick dashed curves correspond to the contributions
at the smallest distance of a given distance range, whereas the thin solid curves
show this contribution at the largest distance. For example, in the upper left
panel the dashed curve shows systematic variations versus  longitude in proper 
motions expected for a cluster at 0.085 kpc from the Sun, whereas the solid curve
shows this variations for a cluster at 0.5 kpc.}
\label{gldist_fig}
\end{figure*}

Fig.~\ref{gldist_fig} shows the distribution of kinematic data of open clusters
versus Galactic longitude for different ranges of distances as indicated at
the top of the figure. The curves mark systematic contributions of Solar motion and
Galactic rotation to the observed velocities of open clusters which are derived
with the parameters in Table~\ref{kinem_tab}. Since this systematic velocity component 
is a function of distances, we show its contribution with the dashed and solid curves
for the minimum and maximum distances in a given distance range, respectively.
The $rms$ errors for
proper motions and radial velocities are taken from the \clucat, 
the errors of tangential velocities are derived as:
\begin{equation}
\veps^2_T = k^2\,d^2\,\mu^2\,(\delta^2_\mu+ \delta^2_d), \label{errt_eqn}
\end{equation}
where $\delta_\mu=\veps_\mu/\mu,\,\delta_d=\veps_d/d$ are the relative errors of a
proper motion and distance. The relative error of a proper motion $\delta_\mu$
is computed from the individual data of clusters, whereas $\delta_d$ is assumed to be 
$15\%$ of a distance (see Paper~II).

In an ideal case, the observed velocities should be located between the dashed and solid
curves in each panel of Fig.~\ref{gldist_fig}.
Deviations significantly larger than the individual $rms$ errors indicate 
peculiarities in cluster motions. For example, a few clusters at longitudes
44...124$^\circ$ and distances 0.5...1.5~kpc show unusually large 
proper motions and tangential velocities. 
We found that all of them are older than $\log t = 8.6$. With the exception of NGC~6991
their radial velocities are unknown, so we cannot determine the complete 
space velocity vector. Fortunately, their location close to $l=90^\circ$ allows us
to estimate the $U$-components, and we found that these clusters move towards the
Galactic anticentre: $U$-velocities corrected for the Galactic rotation are from $-30$ to $-50$ km/s.
Direct calculations of the velocity
vector for NGC~6991 give a similar radial component $U=-58$ km/s. The age of the clusters
as well as their motion towards the Galactic anticentre remind us of the Hyades
($\log t = 8.90$, $U = -43.5$ km/s). We consider this cluster group in \S~\ref{inhomo_sec}
in more detail.

Oort's constants of Galactic rotation  are related to the distance of a cluster 
from the Galactic centre $R_G$, and to the angular velocity $\Omega$ of the disk as: 
\begin{eqnarray}
A &=& -\frac{R_{G,0}}{2}\left(\frac{d\,\Omega}{d\,R_G} \right)_0, \\
B &=& A - \Omega_0 \nonumber 
\label{oort_eqn}
\end{eqnarray}
Assuming the galactocentric distance of the Sun $R_{G,0}$ = 8.5 kpc, we obtain the
angular velocity of the Galactic rotation $\Omega_0 = A - B = 27.5 \pm 1.3$ km/s/kpc
and the corresponding rotation velocity of the system of open clusters as 233.8 km/s.

This result is in a good agreement with the rotation parameters derived from Hipparcos
data over a similar range of ages (less than 1 Gys). For example, Feast \& 
Whitelock~(\cite{feast}) obtained $(A,B)=(14.82\pm0.84, -12.37\pm0.64)$ km/s/kpc
from kinematic data on 220 Galactic cepheids located mainly within 4 kpc from the Sun.
Using data on about 2000 O--B stars, Torra et al.~(\cite{tfg00})
found $(A,B)= (11.8\pm0.4, -12.3\pm0.4)$ km/s/kpc  for $d\le 2$ kpc. From data
on about 1820 O--B5 stars within 1 kpc from the Sun, Branham~(\cite{bra02}) derived
$(A,B)= (+14.9\pm0.8, -15.5\pm0.7)$ km/s/kpc.

Also, the parameters of Solar motion derived with open clusters agree quite well
with previous results for the moderately young disk. Based on a kinematically unbiased
sample from the Hipparcos catalogue in the Solar velocity, Dehnen \& Binney~(\cite{debi98}) 
determined the Solar motion with respect to the LSR as 
$(U,V,W)_{\odot}=(10.00\pm0.36,5.25\pm0.62,7.17\pm0.38)$ km/s. Whereas the
$U$ and $W$-components are in a perfect agreement with the results derived
with open clusters, the rotation component is about two times smaller
than in Table~\ref{kinem_tab}. The difference is mainly caused by the
different evolution parameters of the samples used. The sample of
Dehnen \& Binney contains Main Sequence stars from the Solar neighbourhood
with $(B - V ) \lesssim 1.2$ i.e., stars of different ages, from a few million years
up to 10 Gyr. As we already mentioned before, our cluster sample represents
the population of the moderately young disk. 
(see also \S~\ref{ltime_sec}). Indeed, comparing our results with that of
Torra et al.~(\cite{tfg00}) who found $(U,V,W)_{\odot}=(11.00\pm0.2, 12.9\pm0.2,6.8\pm0.1)$ km/s
with OB stars, we see a good agreement for all components of the Solar motion.
We conclude that both systems of young stars and of open clusters move
similarly.

Since the age for each open cluster of our sample was determined by the same method,
we can check whether real trends in kinematic parameters exist, depending on ages. The most
interesting trend is a dependence of the cosmic (or true) velocity dispersion on the
cluster age. Due to the inhomogeneity of the cluster $RV$s, it is rather difficult to
estimate a realistic contribution of the $rms$ errors in the measurements to the observed
dispersion of space velocities.
Therefore, we prefer to use the dispersion of tangential velocities
which can be properly corrected for the $rms$ error of observations.
Let $\veps_T$ be the mean error of tangential velocities of a given cluster sample,
then
\begin{equation}
\tilde{\sigma}^2_T =  \sigma_T^2 - \veps_T^2 \label{dispt_eqn} \\
\end{equation}
where $\tilde{\sigma}_T$ is the wanted cosmic dispersion, 
and $\sigma_T$ is the dispersion of observed tangential velocities corrected 
for the Solar motion and Galactic rotation. We consider all clusters within $d^c_{xy}$,
and with $\log t$ from 6.7 to 9.5 binned in 10 overlapping groups of ages. The
resulting relation is well approximated by the following equations:
\begin{eqnarray}\label{resel_eqn}
\tilde{\sigma}_{T_l} &=& (1.44\pm0.45)\,\log t_6 +(3.93\pm0.84),\\ \nonumber
\tilde{\sigma}_{T_b} &=& (1.33\pm0.17)\,\log t _6+(1.65\pm0.33)  
\end{eqnarray}
where $t_6$ is the cluster age in units of Myr. 
For a given $\log t$, $(\tilde{\sigma}_{T_l},\tilde{\sigma}_{T_b})$ represent 
the expected dispersions of tangential velocities of clusters of this age and describe
the ellipse of the velocity dispersion, or simply the velocity ellipse.
Since the $T_l$ components might be biased by residual rotations, we use the
second equation of (\ref{resel_eqn}) for estimating the ellipsoid of the cosmic
dispersion. Moreover, for a flat system of open clusters the $T_b$ component
does not differ significantly from $W$.  
Using the ratio $(\sigma_U:\sigma_V: \sigma_W)$ in 
Table~\ref{kinem_tab}, we derive the ellipsoid of the cosmic dispersion
as $(7.2:4.1:2.6)$ km/s for young open clusters with $\log t=6.7$,  and
$(17.5:10.0:6.3)$ km/s for old clusters with $\log t=9.5$.

The derived relations were compared with recent findings. 
Dehnen \& Binney~(\cite{debi98}) determined dispersions for MS stars in
different colour ranges. Their youngest group was selected in the  $(B-V)$-range
from $-0.238$ to 0.139. According to our estimates, the average $\log t$ of
these stars should be somewhere between 8.2 and 8.8. Then, from
eq.~(\ref{resel_eqn}) we derive $\tilde{\sigma}_W=4.6...5.4$ km/s. The
corresponding value of  Dehnen \& Binney~(\cite{debi98}) is $5.46\pm0.61$ km/s
and fits well to our estimates. According to Torra et al.~(\cite{tfg00}), the
velocity dispersion of OB stars increases from 4.3 km/s for about 15 Myr old
stars to 5.5 km/s for  stars older than 120 Myr. These results are comparable to
ours, although they are larger by about 20\%, than those derived from
eq.~(\ref{resel_eqn}).

\section{The local density inhomogeneities}\label{inhomo_sec}

In this section we discuss the density enhancement of open clusters found
in the spatial distribution in the Solar neighbourhood illustrated
in Fig.~\ref{ddist_fig}. The density peaks are detected neither only in one
age group (this would be the case if a peak is a signature of one physical group
of clusters), nor in all of them (which would indicate a general property
of the cluster distribution). On the contrary, the clusters from groups T2 and T4 show
distributions which are compatible with a constant density of clusters,
whereas the samples T1 and T3 have significant maxima within the 
completeness area. Moreover, the samples T1 and T3 display differences in their
density distribution. In addition to the prominent peak at about 0.4~kpc, the sample 
T1 shows a secondary density excess of lesser significance at lower distances which
can be interpreted as evidence for a prolate structure. The sample T3 demonstrates
only one, but a broader density excess. In the following we consider the properties of each
sample separately, and try to answer the question: are these density peaks 
manifestations of independent ``clusterings'' of open clusters? If so, to which degree are
they separated from each other, and from the general cluster substratum?

Hereafter, we call such cluster groups ``open cluster complexes'' (OCCs), in
analogy to large star forming complexes which include associations and young
clusters  (see e.g., Janes et al.~\cite{janea88},
Efremov~\cite{efr95}). We would like to stress that unlike star forming
complexes, which are considered as aggregates of objects related to star
formation, the OCCs have been primarily identified within  a sample of classical
open clusters. Of course, some of them could (but need not) coincide with star
forming regions. In the following, we refer to clusters, which are not found to
be members of OCCs,   as ``field'' clusters of the Galactic disk.

\begin{figure}[tb] \resizebox{\hsize}{!}
{\includegraphics[bb=65 60 560 675,angle=270,clip=]{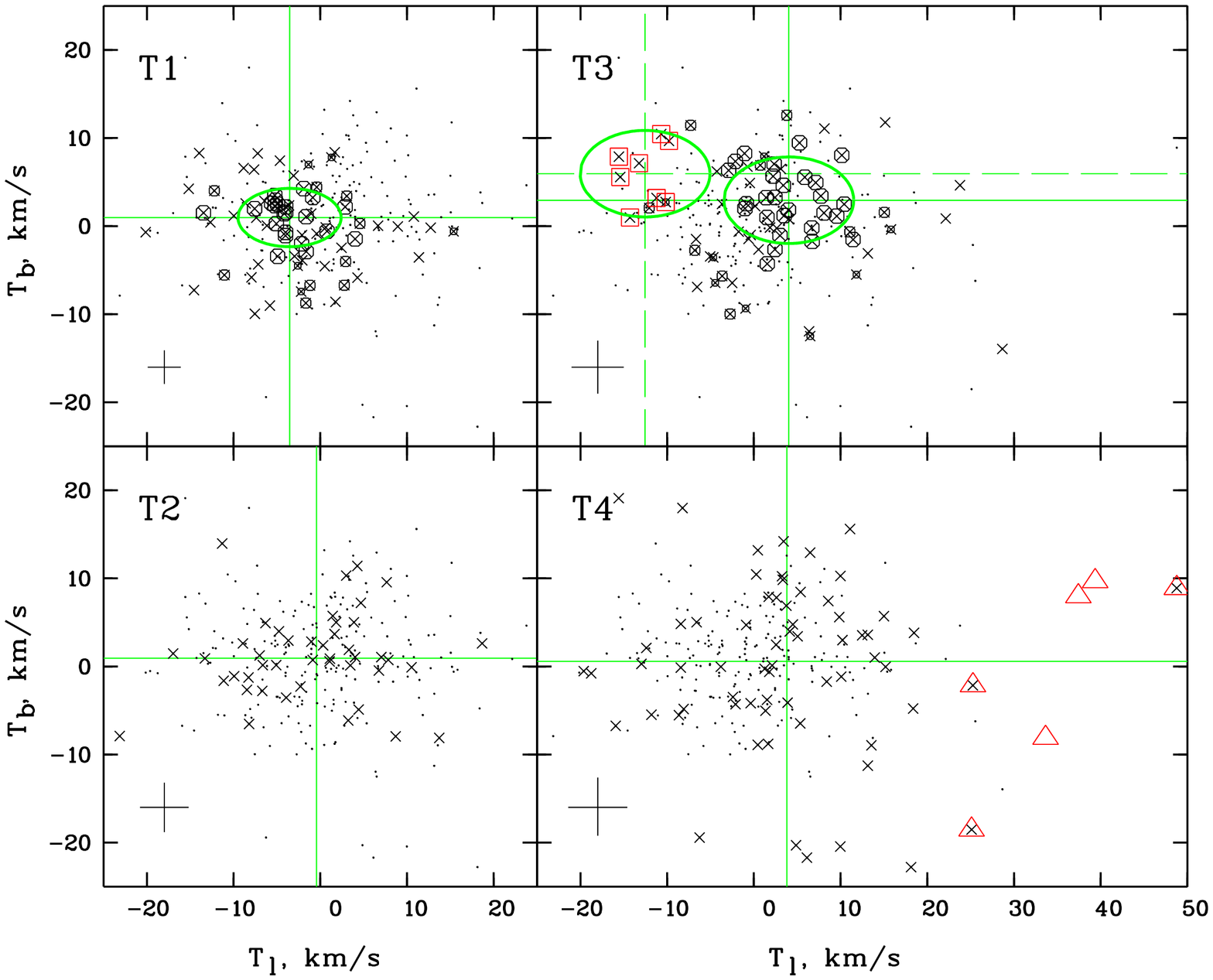}}
\caption{Vector point diagrams of tangential velocities of open clusters in
four age samples (from T1 to T4). The velocities are corrected for Solar
motion and differential rotation of the Galaxy. Dots
are clusters (independent of age) within the completeness area. The 
crosses mark clusters of a given age sample. The circles and squares are
clusters located within the areas of the density excesses (samples T1 and T3).
Their sizes indicate kinematic membership probability to belong to an OCC
correspondingly, the largest symbols indicate $1\sigma$ members, the intermediate
circles are $2\sigma$ members, and the smallest circles are $3\sigma$ members.
The squares (in T3) mark the members of the Perseus-Auriga group 
(see \S~\ref{g3_sec}). The triangles (in T4) indicate possible members of
the Hyades cluster group (see \S~\ref{kin_sec} and \S~\ref{g4_sec}). The
plusses in the bottom-left corners in each panel give the mean errors of 
tangential velocities of open clusters. The (green) cross-hair indicates 
the average tangential motion of each group, whereas the ellipses show the
velocity ellipses of the complexes.
}\label{gkin_fig}  
\end{figure}

\subsection{Tangential velocity and membership in complexes}\label{memb_sec}

In order to get additional pros and/or contras for the physical nature of OCCs,
we consider the distribution of kinematic parameters of open clusters. Since
radial velocities have not been measured for all clusters, we restrict ourselves
to the study of two-dimensional kinematics only. As the clusters are located 
at different distances from the Sun, we use tangential velocities instead 
of proper motions. 

In Fig.~\ref{gkin_fig} we compare the distributions of clusters of different ages
in the $(T_b,T_l)$ diagrams (i.e., the vector point diagram of tangential velocities,
the VPD).
The $T_b,T_l$ velocities are corrected for Solar motion and differential
rotation of the Galaxy. Here we consider only clusters located within the completeness
area ($d_{xy} \le 0.85$~kpc) and compare their distribution with distribution of
clusters in the enhancement areas (in the samples T1 and T3).

A proper selection of kinematic members requires objective criteria taking
into account peculiarities of the kinematic data used. Although we may apply
a technique of probability determination normally used for the selection of
members of a cluster, we must introduce some principal modifications. Determining
the members from the proper motion distribution within a small cluster area, we may assume
that proper motions of cluster members differ from the average only by
their $rms$ errors, and generally, the internal motion is negligibly small
compared to uncertainties of proper motions. This assumption is not always true
if we consider the common motion of clusters belonging to an OCC. According to 
eq.~(\ref{resel_eqn}), we would expect a cosmic dispersion comparable to the 
$rms$ errors of the tangential velocities. Due to the definition of the
tangential velocities, even clusters with similar space velocities should show
different tangential velocities as a function of their galactic coordinates.
This effect will introduce an additional spread of data points in the VPD.
Since the clusters of our sample represent a flat system, this spread should
be more significant in $T_l$ than in $T_b$. Let us assume an ideal (i.e., without
internal rotation and velocity dispersion) stream of clusters located in the
Galactic plane and well distributed around the Sun. Then, the dependence of $T_l$ on
galactic longitude $l$ for each cluster has a sine function:
\begin{equation}
T_l = -A_{T_l}\sin (l-l_0)
                       \label{tansin_eqn}
\end{equation}
where $l_0$ is the apex direction of the stream and $A_{T_l}$ is the maximum
contribution of the velocity component $T_l$ to the space velocity of the
stream. An additional rotation will simply make the mean  tangential velocity
$\olin{T}_l$ of a cluster group different from zero.
It is self-evident that the spread of tangential velocities in the VPD
will be smaller if OCC members are located at similar galactic longitudes as
well as when they have low common velocities with respect to the centroid of 
open clusters. However, in reality the dependence will be more or less disturbed by 
a spread in the location of clusters along the Z-axis, by possible
internal motions, uncertainties of kinematic data, and so on. Below we 
discuss the kinematic behaviour of the complexes in further detail.
 
Due to the reasons described above, we cannot consider the differences
of cluster motions with respect to their group average to be a measure of the 
membership probability
as we did in the case of stars in a cluster. In the case of OCCs, we must consider
residual motions of individual clusters with respect to 
the velocity ellipse defined by a realistic dispersion of tangential velocities.

The membership probability $P_t$ of a cluster to belong to a OCC is computed as
\begin{equation}
P_t =
\exp\left\{-\frac{1}{2}\left[\left(\frac{\delta T^i_l}{\veps^i_{T_l}}\right)^2 +
                       \left(\frac{\delta T^i_b}{\veps^i_{T_b}}\right)^2 
                       \right]\right\}. 
                       \label{tanprob_eqn}
\end{equation}                         
Here $\veps^i_T$ are $rms$ errors of tangential velocities of the $i$th cluster, and
$\delta T^i$ are residual motions of the cluster with respect to the
velocity ellipse of the OCC considered.
Depending on the residual motions $\delta T^i$, we classify
the membership probability as high ($\delta T^i/\veps^i_{T}<1,\,
P_t>$ 61\%), moderate ($1\le\delta T^i/\veps^i_{T}< 2,\,P_t=$ 14\% - 61\%), and low
($2\le\delta T^i/\veps^i_{T}<3,\,P_t>$ 1\% - 14\%). Traditionally, we call such objects
$1\sigma, 2\sigma$, and $3\sigma$ members (see Paper~I), respectively.
Note that all clusters within the velocity ellipse
have the highest membership probability ($P_t=100\%$).

The parameters of velocity ellipses were chosen in accordance with
equations~(\ref{resel_eqn})
derived in \S~\ref{kin_sec}. The centres of the distribution and membership probabilities
were computed in an iterative procedure. As a first approximation of
the centres, we used a simple average of the tangential velocities
of all clusters at a given age and we derived membership probabilities by
eq.~(\ref{tanprob_eqn}). After that, we selected the $1\sigma$-members
to derive the next iteration of the distribution centre. The iterations
were stopped when the list of the $1\sigma$-members did not change in subsequent
iterations. The results are shown in Fig.~\ref{gkin_fig} and Table~\ref{grpar_tab}.

\begin{figure*}[tb] \resizebox{\hsize}{!}
{\includegraphics[bb=50 210 540 650]{3764f07lr.ps.gz}}
\caption{Spatial distribution of the OCC~1 clusters in 3D-space. The
proper coordinate system $X',Y',Z$ of the OCC~1 is described in the text. The
coordinate axes $X$ and $Y$ of the Galactic rectangular system are shown with
dotted lines. Two circles centered at the origin of the Galactic coordinate
system indicate the outer boundary of the area of the enhanced density at 0.5~kpc
and the completeness area at 0.85~kpc. The solid ellipse outlines the OCC~1 members
selected as described in the text. The dashed ellipse represents the Olano~(\cite{ola82})
model of gas in Gould's Belt, given here for comparison. The small dots are all the
clusters from our sample, the crosses mark T1 clusters, and the filled circles are for
member candidates of the OCC~1. The small (cyan) circles are $3\sigma$ members,
the dotted circles (green) of medium size show $2\sigma$ members, and
the largest dotted circles  (light yellow) with a (green) core indicate $1\sigma$
members. The grey colour contours give the projected density distribution of T1 clusters,
derived separately in each panel. The density distribution was computed
in windows of $50\times50$ pc$^2$ and smoothed with a 5-point rectangular filter.
The rules to the right indicate the corresponding density scales where the $(X',Z)$ and $(Z,Y')$
planes share the same grey colour scale. The straight line in the panel $(Z,Y')$ is
the regression line computed with $1\sigma$ clusters. The resulting inclination
$i$ is given in the left upper corner of $(Z,Y')$ plane.
}\label{g1_xyz_fig} 
\end{figure*}

\subsection{OCC~1: a complex of clusters associated with Gould's Belt}

For the sample T1, we found 37 clusters within $d_{xy}=0.5$ kpc, with 28 of
them located within the main maximum of the density distribution at
$d_{xy}=0.325\ldots0.5$ kpc. Recomputing the surface density distribution
in terms of numbers of clusters, we infer that 16$\pm$4 clusters should 
belong to the field, whereas 20$\pm$4 clusters are responsible for the
density excess. Thus, we may expect about 20 members in a potential OCC.
About the same number of clusters is also obtained from the statistics of
kinematic membership. Table~\ref{grpar_tab} shows that within 0.5 kpc 
from the Sun, we found N = 23 clusters which are $1\sigma$ members from the
distribution of their tangential velocities. We conclude that the $1\sigma$ threshold
for kinematic probability ($P_t=61$\%) separates well the complex members 
from the field.

The distribution of the OCC~1 candidates in 3D-space is shown in
Fig.~\ref{g1_xyz_fig}. For convenience,  we present the results in a coordinate
system related to this cluster group. The coordinate system $X',Y',Z$ is
rotated in the Galactic plane to such that the $X'$-axis corresponds to the line of nodes
of the OCC~1 plane. The centre of the system is
shifted to the apparent centre of the group by $\Delta X=-0.078$ kpc and $\Delta Y=-0.053$ kpc.

According to Fig.~\ref{g1_xyz_fig}, the OCC~1 clusters constitute an elongated configuration
immersing below the galactic plane in the region of the Orion association by about 200 pc,
and then rising above the galactic plane by about 100 pc at the opposite 
direction in the IC~4665 region. The apparent centre of the complex is located close to 
the Sco-Cen group with the clusters Mamajek~1 and Platais~8. On average, the OCC~1
resides below the Galactic plane ($\olin{Z}=-46$ pc for the $1\sigma$ members).
The ascending node of the OCC~1 plane is located near the cluster Alessi~5 at 
$(l,b)=(288\fdg08,-2\fdg05)$, whereas the descending node is somewhere near ASCC~127
at $(l,b) =(112\fdg26,4\fdg13)$. This finding is well compatible with the results by
Torra et al.~(\cite{tfg00}) who derived the longitude of the ascending node 
of Gould's Belt as $l=275\ldots295^\circ$ from the Hipparcos data of OB field stars.
It seems that the density excess of T1 clusters can be associated with the
Gould's Belt complex which is usually believed to contain early-type field stars, associations,
or interstellar gas (e.g., see a review by P\"oppel~\cite{pop97}).

The OCC~1 has a flat structure, but is not strictly planar. In the $(Z,Y')$-plane it
shows a significant inclination (as the regression solution shows) of $19^\circ$
with respect to the Galactic plane. Torra et al.~(\cite{tfg00}) found an inclination of 
$i=16\ldots22^\circ$ from the data on field stars.
In the $(Z,X')$-plane, the complex presents a less pronounced structure.
It is a rather smooth transition from the lowest location near NGC~1980 to Stephenson~1
via Vela~OB1 and Mamajek~1, and then a steeper rising from IC~348 to IC~4665
via ASCC~127.

\begin{table}
\tabcolsep 3pt
\caption[]{Candidates of the OCC~1 (the Gould's
Belt complex)}
\label{GBlist_tab}
\begin{tabular}{rlrrrrr}
\hline

 COCD& Name          &$   X  $&$   Y  $&$   Z  $&$\log t$&$P_t$ \\
     &               &   kpc  &   kpc  &   kpc  & yrs &  \%  \\ 
 \hline               $      $ $      $ $      $
     &               &$      $&$      $&$      $&    &       \\
  41 &Stock 23       &$-0.291$&$ 0.244$&$ 0.014$&7.51&  25 \\
  42 &Melotte 20     &$-0.159$&$ 0.102$&$-0.021$&7.55&   2 \\
  46 &IC 348         &$-0.354$&$ 0.125$&$-0.120$&7.79&  84 \\
  68 &Collinder 65   &$-0.301$&$-0.047$&$-0.058$&7.41& 100 \\
  72 &Collinder 69   &$-0.413$&$-0.113$&$-0.092$&6.76& 100 \\
  73 &NGC 1981       &$-0.334$&$-0.178$&$-0.130$&7.50& 100 \\
  74 &NGC 1976       &$-0.329$&$-0.183$&$-0.132$&7.71& 100 \\
  75 &NGC 1977       &$-0.415$&$-0.225$&$-0.164$&7.08&  40 \\
  76 &NGC 1980       &$-0.451$&$-0.255$&$-0.184$&6.67& 100 \\
  77 &Collinder 70   &$-0.338$&$-0.158$&$-0.117$&6.71& 100 \\
  80 &Sigma Ori      &$-0.340$&$-0.172$&$-0.119$&6.82&   5 \\
  91 &Platais 6      &$-0.313$&$-0.148$&$-0.038$&7.79&   1 \\
  95 &NGC 2232       &$-0.266$&$-0.183$&$-0.042$&7.49& 100 \\
 126 &Collinder 132  &$-0.183$&$-0.362$&$-0.066$&7.51& 100 \\
 133 &Collinder 135  &$-0.112$&$-0.292$&$-0.062$&7.54&  11 \\
 136 &Collinder 140  &$-0.168$&$-0.361$&$-0.055$&7.57&  15 \\
 159 &NGC 2451A      &$-0.056$&$-0.178$&$-0.025$&7.76&  36 \\
 162 &NGC 2451B      &$-0.132$&$-0.406$&$-0.050$&7.88&  19 \\
 182 &Vel OB2        &$-0.048$&$-0.404$&$-0.057$&7.26& 100 \\
 183 &NGC 2547       &$-0.044$&$-0.450$&$-0.068$&7.70&  23 \\
 190 &vdB-Hagen 23   &$-0.120$&$-0.420$&$-0.008$&7.14& 100 \\
 202 &IC 2391        &$ 0.001$&$-0.175$&$-0.021$&7.88&  73 \\
 204 &Mamajek 1      &$ 0.037$&$-0.090$&$-0.039$&6.90& 100 \\
 210 &Trumpler 10    &$-0.052$&$-0.414$&$ 0.005$&7.38& 100 \\
 216 &Platais 8      &$ 0.020$&$-0.147$&$-0.020$&7.75&  27 \\
 259 &IC 2602        &$ 0.053$&$-0.150$&$-0.014$&7.83&  13 \\
 261 &Alessi 5       &$ 0.123$&$-0.378$&$-0.014$&7.71& 100 \\
 412 &IC 4665        &$ 0.290$&$ 0.171$&$ 0.103$&7.63& 100 \\
 456 &Stephenson 1   &$ 0.142$&$ 0.331$&$ 0.099$&7.69& 100 \\
 479 &Roslund 5      &$ 0.133$&$ 0.396$&$ 0.002$&7.77&  28 \\
1016 &ASCC 16        &$-0.408$&$-0.156$&$-0.145$&6.93& 100 \\
1018 &ASCC 18        &$-0.439$&$-0.178$&$-0.159$&7.12& 100 \\
1019 &ASCC 19        &$-0.299$&$-0.139$&$-0.117$&7.64& 100 \\
1020 &ASCC 20        &$-0.399$&$-0.158$&$-0.135$&7.35& 100 \\
1021 &ASCC 21        &$-0.451$&$-0.163$&$-0.142$&7.11& 100 \\
1024 &ASCC 24        &$-0.318$&$-0.236$&$-0.057$&6.96&   0 \\
1089 &ASCC 89        &$ 0.415$&$-0.249$&$-0.126$&7.03&  42 \\
1127 &ASCC 127       &$-0.132$&$ 0.323$&$ 0.025$&7.82&  68 \\
\hline                
\end{tabular}         
\end{table}           

In Table~\ref{GBlist_tab} we list all T1 clusters located within $d_{xy}=0.5$
kpc from the Sun. Due to its high membership probability ($P_t$ = 100\%), we
extended this  list by one Orion cluster (NGC~1980) residing at $d_{xy}=518$ pc.
For each cluster in  Table~\ref{GBlist_tab}, we provide the \clucat  number, the
cluster designation, the $(X,Y,Z)$ coordinates, the cluster age, and the
kinematic membership probability $P_t$ for belonging to the Gould's Belt
complex.

As described above, clusters located within the velocity ellipse
got the highest membership probability $P_t$  = 100\%. On the other hand, some
clusters  like e.g. Melotte~20 ($\alpha$~Per) which were believed to be members
of the Gould's Belt system  (e.g., see Olano~\cite{ola82}) have a low kinematic
probability.

Fortunately, all the $1\sigma$ members of OCC~1 have full space velocity vectors.
Therefore, we can analyse the spatial motion of this complex
in more detail. On average, OCC~1 rotates around the Galactic centre at about the
same velocity as the centroid of the open cluster system (the relative velocity is
$\olin{V}=-0.2\pm1.1$ km/s).
The average tangential velocity of the most probable members of the OCC~1
is $\olin{T}_l=-3.6\pm1.4$ km/s i.e., it differs significantly from zero.
Therefore, we may conclude an internal rotation of the OCC~1 complex around
its centre near IC~2391 and Mamajek~1, and in the direction
of  Galactic rotation. Excluding these two clusters which do not contribute to
the determination of the internal rotation, we derive $\olin{T}_l=-2.9\pm0.7$ km/s.
At an average distance of 0.39 kpc from the OCC~1 centre, one obtains the corresponding
angular velocity as $\omega = -8.6\pm2.2$ km/s/kpc. This is considerably
lower (in absolute units) than the value $\omega \approx -20$ km/s/kpc derived by 
Bobylev~(\cite{bobyl}) from the kinematics of Hipparcos stars proposed to be 
members of associations and open clusters.

\begin{figure*} 
\resizebox{\hsize}{!}
{\includegraphics[bb=50 210 540 650]{3764f08lr.ps.gz}}
\caption{Spatial distribution of T3 clusters in 3D-space. The coordinates are
given in the Galactic cartesian system.
The larger circles indicate possible members of OCC2, the diamonds mark
members of the Perseus-Auriga group (in order to avoid an overcrowding in 
the $X,Z$ and $Y,Z$ panels,
they are shown together with the corresponding boundary
circle only in the $X,Y$ panel). The dashed circle gives the inner edge of the
enhancement area.
The other designations are the same as in
Fig.~\ref{g1_xyz_fig}.
}\label{g3_xyz_fig} 
\end{figure*}

The expansion velocity of the OCC~1 computed as a residual motion along 
a radius-vector projected onto the $XY$ plane and with origin in 
the OCC~1 centre is $V_r=2.1\pm1.1$ km/s. This seems to be insignificant.
The latter conclusion confirms
the results of Torra et al.~(\cite{tfg00}) who found an expansion velocity of 
$1.5\pm1$ km/s for young field stars at $d>0.25$ kpc.

The total velocity dispersion $\sigma_{UVW}$ for the 23 members of the OCC~1 
\[
\sigma^2_{UVW}=(\sigma^2_U+\sigma^2_V+\sigma^2_W)/3
\]
is found to be $4.8\pm0.7$ km/s.
This is significantly less than the total dispersion of $7.6\pm0.7$ km/s
computed from all 69 T1 clusters  in the completeness area.
This difference gives an additional confirmation for the distinction of
the complex from the general field.

\subsection{OCC~2 and the Perseus-Auriga group: two in one}\label{g3_sec}

According to the cluster counts within $d_{xy}=0.65$ kpc, the sample T3 contains 53 clusters,
with 48 of them in the location of the main density maximum (i.e., at $d_{xy}=0.325...0.65$ kpc).
The lower density of T3 clusters at $d_{xy}<0.325$ kpc from the Sun can well be seen
in Figs.~\ref{ddist_fig} and \ref{g3_xyz_fig}. 
Recomputing the surface density distribution in terms of numbers of clusters, we
infer that 33$\pm$6 of them belong to field clusters and 20$\pm$6 form the density excess.
The number of kinematic candidates of OCC~2 (i.e., $1\sigma$ members) is 27 (see Table~\ref{grpar_tab}).
Although a difference of 7 clusters is still within the statistical uncertainty,
the disagreement between both estimates is larger than in the OCC~1 case. The average
density used for computing the number of field clusters is determined from the
distribution of T3 clusters up to $d_{xy}=0.85$ kpc, and it is slightly higher
than the density in the inner part of $d_{xy}<0.325$ kpc from the Sun (Fig.~\ref{ddist_fig}).
Assuming the density in the Solar neighbourhood as being representative for the T3 group,
we obtain 20 field clusters and 33 OCC members forming the enhancement, and the
excess would be extended somewhat outward of 0.65 kpc.
Below, we show that, indeed, we found another kinematic group, the Perseus-Auriga group (see
Fig.~\ref{g3_xyz_fig} and Table~\ref{grpar_tab}), among the T3 clusters. The complex
is very compact in space as well as in the VPD and consists of 8 clusters including
one cluster at $d_{xy}=0.77$ kpc i.e., outside the main density excess.
Consequently, the final census of the T3 sample within $d_{xy}=0.65$ kpc is
27 OCC~2 members, 7 members of the Perseus-Auriga group and 19 field clusters. This result coincides well with
the number of 20 field clusters estimated before. Although OCC~2 and
the Perseus-Auriga group are of the same ages, they show different kinematics and
spatial distribution. Therefore we consider these two groups as separate entities.

\begin{table}[t]
\tabcolsep 4pt
\caption[]{Candidates of the OCC~2 }
\label{g3list_tab}
\begin{tabular}{rlrrrrr}
\hline
 COCD& Name        &  $X$ & $Y$  & $Z$  &log $t$&$P_t$ \\
     &             &  kpc &  kpc &  kpc & yrs &  \%    \\ 
 \hline                                               
   3&*Blanco 1     &$ 0.050$&$ 0.013$&$-0.264$&8.32&    2  \\
  15&*Platais 2    &$-0.107$&$ 0.136$&$-0.102$&8.54&  100  \\
  36&*NGC 1039     &$-0.387$&$ 0.285$&$-0.134$&8.42&    4  \\
  89 &NGC 2184     &$-0.537$&$-0.328$&$-0.117$&8.37&    1  \\
  96 &Collinder 95 &$-0.516$&$-0.206$&$ 0.000$&8.36&  100  \\
  97 &Collinder 97 &$-0.452$&$-0.214$&$-0.015$&8.32&    8  \\
 141 &NGC 2396     &$-0.397$&$-0.433$&$ 0.027$&8.52&  100  \\
 144 &NGC 2413     &$-0.287$&$-0.333$&$ 0.024$&8.46&  100  \\
 151 &Ruprecht 27  &$-0.266$&$-0.492$&$-0.025$&8.41&  100  \\
 161 &Ruprecht 31  &$-0.163$&$-0.449$&$-0.050$&8.56&  100  \\
 168 &ESO 123-26   &$ 0.029$&$-0.527$&$-0.154$&8.33&   63  \\
 200 &Ruprecht 65  &$-0.060$&$-0.496$&$-0.014$&8.57&  100  \\
 223 &Turner 5     &$-0.040$&$-0.391$&$ 0.077$&8.49&    0  \\
 241 &NGC 3228     &$ 0.092$&$-0.489$&$ 0.040$&8.43&   19  \\
 245 &Loden 143    &$ 0.158$&$-0.579$&$-0.009$&8.45&   59  \\
 246 &Loden 89     &$ 0.094$&$-0.368$&$ 0.006$&8.47&  100  \\
 247 &Loden 59     &$ 0.147$&$-0.632$&$ 0.037$&8.45&   61  \\
 276 &NGC 3532     &$ 0.167$&$-0.468$&$ 0.012$&8.45&   65  \\
 333 &Loden 915    &$ 0.310$&$-0.391$&$ 0.028$&8.44&   24  \\
 340 &Loden 1171   &$ 0.332$&$-0.372$&$ 0.029$&8.45&   27  \\
 344 &Loden 1194   &$ 0.336$&$-0.370$&$ 0.016$&8.53&   72  \\
 349 &ESO 175-06   &$ 0.386$&$-0.390$&$ 0.038$&8.60&   63  \\
 373 &NGC 6124     &$ 0.441$&$-0.154$&$ 0.049$&8.34&  100  \\ 
 392 &NGC 6281     &$ 0.482$&$-0.105$&$ 0.017$&8.51&  100  \\ 
 410 &Alessi 9     &$ 0.185$&$-0.053$&$-0.031$&8.42&    1  \\ 
 419 &NGC 6469     &$ 0.546$&$ 0.063$&$ 0.019$&8.36&   86  \\ 
 422 &NGC 6494     &$ 0.618$&$ 0.108$&$ 0.031$&8.52&  100  \\ 
 463&*Stock 1      &$ 0.174$&$ 0.304$&$ 0.014$&8.53&   14  \\ 
 469&*NGC 6828     &$ 0.405$&$ 0.432$&$-0.097$&8.56&   41  \\
 489&*Roslund 6    &$ 0.093$&$ 0.440$&$ 0.002$&8.47&    7  \\ 
 496&*NGC 7058     &$-0.020$&$ 0.400$&$ 0.004$&8.35&   21  \\ 
 499&*NGC 7092     &$-0.013$&$ 0.310$&$-0.012$&8.57&    0  \\ 
 514&*Stock 12     &$-0.145$&$ 0.368$&$-0.059$&8.45&  100  \\ 
1034 &ASCC 34      &$-0.474$&$-0.270$&$ 0.067$&8.55&  100  \\ 
1038 &ASCC 38      &$-0.370$&$-0.333$&$ 0.047$&8.60&   16  \\ 
1041 &ASCC 41      &$-0.280$&$-0.229$&$ 0.079$&8.44&    0  \\ 
1051 &ASCC 51      &$ 0.136$&$-0.466$&$-0.122$&8.53&  100  \\ 
1059 &ASCC 59      &$ 0.131$&$-0.534$&$-0.005$&8.60&   70  \\ 
1074 &ASCC 74      &$ 0.343$&$-0.429$&$ 0.034$&8.50&   27  \\ 
1083 &ASCC 83      &$ 0.509$&$-0.317$&$ 0.012$&8.40&  100  \\ 
1101&*ASCC 101     &$ 0.128$&$ 0.318$&$ 0.070$&8.52&  100  \\ 
1106 &ASCC 106     &$ 0.374$&$ 0.318$&$-0.093$&8.42&    1  \\ 
1109&*ASCC 109     &$ 0.151$&$ 0.423$&$ 0.028$&8.31&  100  \\ 
1115&*ASCC 115     &$-0.078$&$ 0.594$&$-0.027$&8.59&  100  \\ 
1123&*ASCC 123     &$-0.064$&$ 0.241$&$-0.017$&8.41&  100  \\ 
1124&*ASCC 124     &$-0.119$&$ 0.576$&$-0.119$&8.48&  100  \\ 
\hline              
\end{tabular}       
\end{table} 
The average age of $1\sigma$ members of OCC~2 is $\log t=8.48$, with a 
standard deviation $\sigma_{\log t}=0.09$.
The distribution of the T3 clusters in 3D-space is shown in Fig.~\ref{g3_xyz_fig}
in the Galactic cartesian coordinate system $XYZ$.  Kinematic members of OCC~2
form a ring-like structure with an outer diameter of about 1.3 kpc, and there are
only a few T3 clusters in its inner part.  The ring is ripped along the main Galactic
meridian, or more exactly along a line at $l\approx15^\circ$, $l\approx195^\circ$,
where no $1\sigma$ members are found. It even seems that $1\sigma$ members form
two subgroups which are separated by this line. For convenience, we call these
subgroups  OCC~2a ($195^\circ < l < 15^\circ$) and OCC~2b ($15^\circ < l < 195^\circ$).
Comparing the visible distribution of the OCC~2 members with the extinction maps, 
one can find an agreement in the locations of the empty region and the Ophiuchus and 
Perseus-Auriga clouds. Thus, we cannot exclude that some OCC~2 members are possibly hidden
by obscuring clouds.

Although less prominent, OCC~2a and OCC~2b differ also in their location in the
$YZ$ plane. OCC~2b shows a dispersed distribution and tends to a mean location
below the Galactic plane, whereas the majority of the OCC~2a members are above
the Galactic plane. In the VPD of tangential velocities, the OCC~2b members are
distributed more compactly around the centre than the members of the OCC~2a
subgroup.

In Table~\ref{g3list_tab} we list all OCC~2 candidates i.e., all T3 clusters
located within $d_{xy}=0.65$ kpc, which are not $1\sigma$ members of
the Perseus-Auriga group. For each cluster, the table contains the \clucat  number,
the cluster designation, $(X,Y,Z)$ coordinates, the cluster age, and membership
probability based on the distribution of tangential velocities in the VPD. The
OCC~2b candidates are marked by asterisks.

\begin{table}
\tabcolsep 6pt
\caption[]{Candidates of the Perseus-Auriga group of open clusters}
\label{occ3_tab}
\begin{tabular}{rlrrrrr}
\hline

 COCD& Name          & $X$    & $Y$   &  $Z$   &$\log t$\\
     &               &  kpc   &  kpc  &  kpc   & yrs   \\ 
 \hline
  37 &NGC~1027       &$-0.553$&$0.539$&$ 0.021$&$8.55$ \\
  40 &Trumpler~3     &$-0.333$&$0.300$&$ 0.035$&$8.34$ \\
  43 &King~6         &$-0.361$&$0.269$&$-0.001$&$8.41$ \\
  52 &NGC~1582       &$-0.560$&$0.212$&$-0.032$&$8.60$ \\
  55 &Alessi~2       &$-0.441$&$0.232$&$ 0.056$&$8.60$ \\
  81 &Stock~10       &$-0.375$&$0.055$&$ 0.023$&$8.42$ \\
1012 &ASCC~12        &$-0.478$&$0.146$&$-0.016$&$8.42$ \\
1023 &ASCC~23        &$-0.567$&$0.126$&$-0.149$&$8.45$ \\
\hline               
\end{tabular}        
\end{table}          

For eight $1\sigma$ members of OCC~2 radial velocities are known, so we
can determine the space velocity vector though, less reliably than for OCC~1
(see Table~\ref{grpar_tab}). According to the rotation velocity around the
Galactic centre, OCC~2 outruns the open cluster centroid with a 
velocity $\olin{V}=3.4\pm1.2$ km/s.
The average tangential velocity in longitude $\olin{T_l}=4.0\pm0.8$ km/s
indicates a significant rotation of OCC~2 with respect to the centroid of open clusters. 
With an average heliocentric distance of OCC~2 of
about 0.5 kpc, we find an angular rotation velocity $\omega=8\pm1.6$ km/s/kpc.

Summarising the main attributes of OCC~2, we conclude that the OCC~2 members
show a number of properties expected for a real physical system. The complex is
limited in space and it shows a significant (by a factor of 2) density enhancement
over the surrounding background. The T3 sample has structural parameters
($Z_0$ and $h_Z$ in Table~\ref{zpar_tab}) which differ from the other samples of clusters,
and for the OCC~2 members these parameters are even different from that of the
T3 field clusters. The OCC~2 members have similar ages and they show  similar
(and relatively compact) kinematics. Therefore, we believe that
OCC~2 with its 27 $1\sigma$-members represents a
real group of clusters formed about 300 Myr ago.

The other group of T3 clusters, which we call the Perseus-Auriga group due to its
special location in space is clearly separated in
Figs.~\ref{gkin_fig} and \ref{g3_xyz_fig}. 
The group has very compact kinematics, the average motion being 
$(\olin{T}_l,\olin{T}_b)=(-12.6\pm0.8,5.9\pm1.2)$ km/s with standard deviation of
$(2.4\pm0.6,3.5\pm0.8)$ km/s. This velocity dispersion is even smaller
than the cosmic dispersion predicted for clusters of corresponding ages
by eq.~(\ref{resel_eqn}) and shown in Fig.~\ref{gkin_fig}. 
However, the compactness of the velocity distribution is not surprising for the
Perseus-Auriga group. The velocity ellipses derived by eq.~(\ref{resel_eqn}) define a
measure of cosmic dispersion for clusters of a given age, provided that the clusters are
distributed all over the sky. Due to the definition of tangential velocities, their
dispersion depends on galactic longitude. For a sample of clusters located within a 
small range of galactic longitudes (like the Perseus-Auriga group), the 
$T_l$-distribution should be
more compact than predicted by eq.~(\ref{resel_eqn}).

Despite of attempts to find more members of the Perseus-Auriga group outside 
the density excess, we could add
only one more cluster. Thus, the group remains very compact in 3D-space.
However, this does not exclude the possibility that there could be more members
since the group is located behind the Perseus-Auriga complex of molecular clouds,
and we see the its members in transparency windows within the clouds.

Only for two clusters of the Perseus-Auriga group radial velocities are available.
Therefore, the accuracy of the space velocity is very poor. The data on the
Perseus-Auriga group is presented in Table~\ref{occ3_tab} and Table~\ref{grpar_tab}.
Since all clusters of the Perseus-Auriga group have tangential velocities distributed within the
corresponding dispersion ellipsoid, their kinematic probability is 100\%.
Therefore, we exclude the probability column in Table~\ref{occ3_tab}.

\subsection{The Hyades moving group of open clusters}\label{g4_sec}
      
According to Fig.~\ref{ddist_fig}, the T4 sample does not show any
peculiarity which would require a dedicated discussion. However, this
sample, also, includes a cluster group which becomes evident due to its
prominent motion detected in \S~\ref{kin_sec} (see Fig.~\ref{gldist_fig}).

We found six open clusters with considerably large tangential velocities. Only
one of them, NGC 6991, has a radial velocity measured, so for this cluster we can determine directly
the complete vector of space velocity corrected for the Galactic rotation
which turns out to be ($U,V,W$) = (--58,--4,2) km/s. The other clusters
are located at longitudes from 44$^\circ$ to 124$^\circ$ and therefore, their $-T_l$
velocities should more or less well coincide with $U$ component, which we have
estimated to be between --30 and --50 km/s. All these clusters are relatively old,
with $\log t$ between 8.57 and 9.16. 

\begin{table}
\tabcolsep 6pt
\caption[]{Candidates of the Hyades moving group of open clusters}
\label{hya_tab}
\begin{tabular}{rlrrrr}
\hline
 COCD& Name          & $X$  & $Y$  & $Z$  &$\log t$\\
     &               &  kpc &  kpc &  kpc & yrs \\ 
 \hline
     &               &      &      &      &     \\
  13 &Alessi~1       &$-0.427$&$ 0.651$&$-0.184$&8.85 \\
 201 &Praesepe       &$-0.142$&$-0.069$&$ 0.100$&8.90 \\
 459 &NGC~6738       &$ 0.500$&$ 0.489$&$ 0.038$&9.16 \\
 461 &NGC~6793       &$ 0.612$&$ 0.912$&$ 0.064$&8.64 \\
 462 &NGC~6800       &$ 0.510$&$ 0.857$&$ 0.069$&8.59 \\
 494 &NGC~6991       &$ 0.032$&$ 0.699$&$ 0.020$&9.11 \\
 502 &NGC~7209       &$-0.118$&$ 1.152$&$-0.150$&8.65 \\
1099 &ASCC~99        &$ 0.267$&$ 0.076$&$-0.039$&8.71 \\
     &Hyades         &$-0.045$&$ 0.000$&$-0.019$&8.90 \\
\hline               
\end{tabular}        
\end{table}          

\begin{figure}
 \resizebox{\hsize}{!}
{\includegraphics[bb=252 50 554 568,angle=270,clip=]{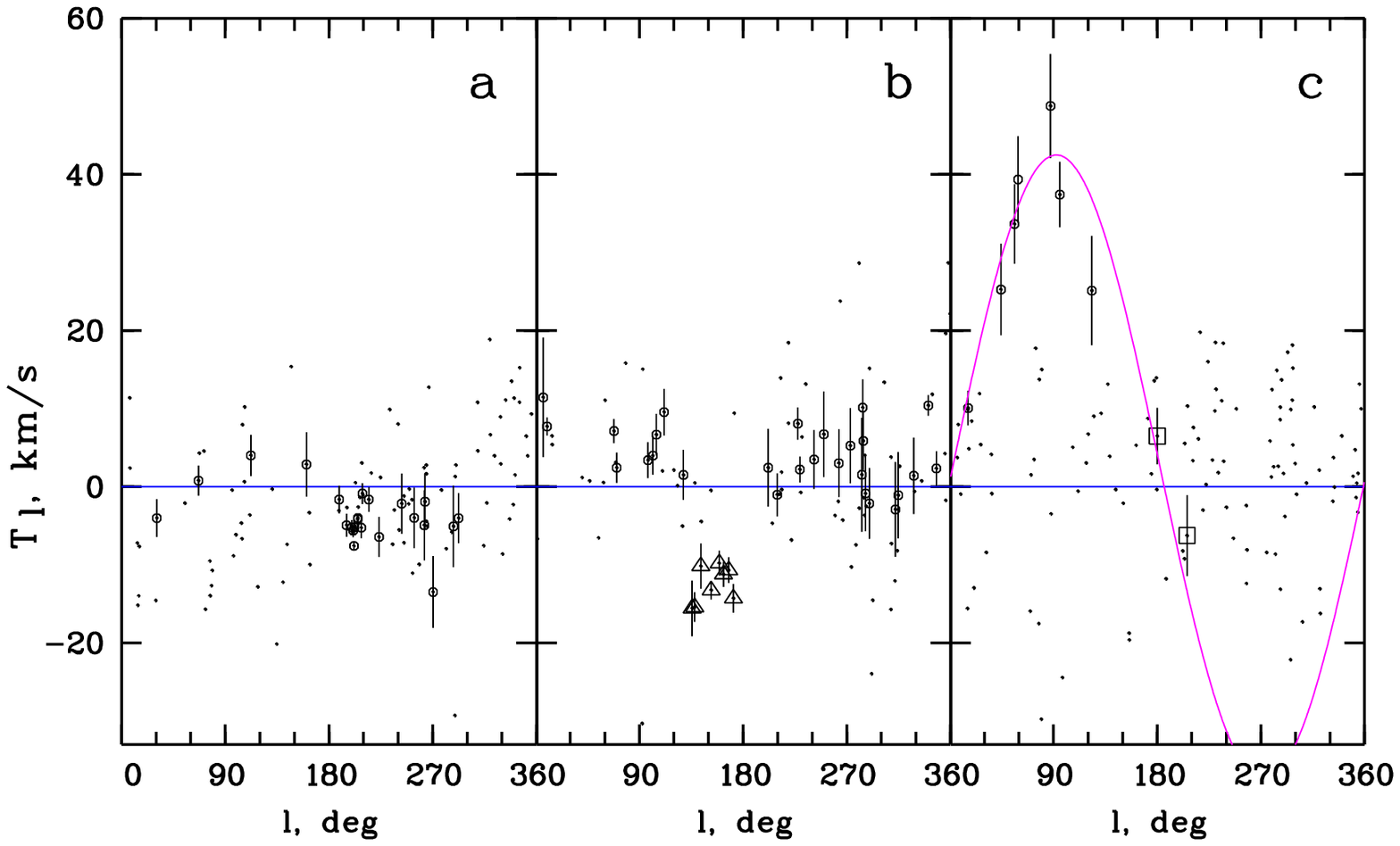}}
\caption{Tangential velocities of clusters versus Galactic longitude. 
The velocities are corrected for differential Galactic rotation and for Solar motion
with respect to the cluster centroid. The points mark clusters of the corresponding
sample. Panel a: the T1 sample, circles are for the OCC~1 members. Panel b:
the T3 sample, circles and triangles indicate OCC~2 and Perseus-Auriga members,
respectively. Panel c: the T4 sample with candidates (circles) of a moving
group. The Hyades and Praesepe are marked by squares. The distribution of
tangential velocities of candidates is fitted by a sine function. 
}
\label{gl_tl4n_fig}
\end{figure}

In order to explain a possible nature of their large tangential velocities,  we
considered distribution of the tangential velocities corrected for the Galactic
rotation and for the Solar motion with respect to the cluster centroid versus
longitude. Since these clusters have large space velocities, there is a good
chance to observe a sine dependence described by eq.~(\ref{tansin_eqn}) if the
clusters belong to a common stream. Fig.~\ref{gl_tl4n_fig} shows this
dependences for the T4 sample (panel c). For comparison, the distribution is
also given for the samples T1 (panel a) and T3 (panel b) where the members of
the complexes OCC~1, OCC~2 and the Perseus-Auriga group are marked, too. 

The visible distribution of tangential velocities of these clusters can be
fitted by a sine function. This indicates a possible stream of
several clusters moving in the direction of the Galactic anticentre with a velocity of about 40~km/s
with respect to the general cluster centroid. We found three other old clusters
with large $U$-motions which fit to the sine function quite well though their
tangential velocities are small (due to their close location to the
centre - anticentre direction). Two of them are well known clusters, the Hyades and Praesepe,
with an age of $\log t$ = 8.9 (both) and $U$-components relative to the
cluster centroid of --34.1 and --30.9 km/s, respectively.
It is possible (but not necessary) that these nine clusters form a common kinematic stream.
The peculiarity of the spatial distribution of the clusters prevents us from 
applying the member selection techniques used for other groups.
For a clear proof radial velocities would be essential. Therefore, we introduce
this moving group only as a working hypothesis to explain the inhomogeneity in peculiar
motion of clusters and call it the Hyades moving group 
after the most famous potential member. This cluster group must not be confused with 
the Hyades supercluster proposed by Eggen~(\cite{eggen58}, \cite{eggen75}) and recently studied
by Famaey et al.~(\cite{fjlm05}) who considered several hundreds of stars in the 
Solar vicinity (up to 500 pc) which have similar 
space velocities as the Hyades. In our case, there are a few old open clusters which are 
located at distances
up to about 1.2~kpc from the Sun but have ages and $U$-velocities comparable with
those of the Hyades.

The list of candidates of the Hyades moving group is given in Table~\ref{hya_tab} and
Table~\ref{grpar_tab}.

\subsection{Summarising cluster complexes}

We classify the complexes OCC~1 and OCC~2 as
real physical formations, the identification of which is supported by a sufficient amount
of data. However, it is possible that not all their members are completely revealed.
For the other two (the Perseus-Auriga and specially, the Hyades group), only fragmentary data are available.
These make the discussion and argumentation of their nature rather difficult.
To make a difference in the identification status, we refer to them as the
groups of clusters. In Table~\ref{grpar_tab} we summarise the parameters of the cluster
complexes and the groups of clusters found in this study. The first column gives
the names introduced in this paper, $N$ is the number of probable members. The
average age with the standard deviation is given in column 3. For OCC~1
which probably shows a real age spread, we provide the age span only. The next
three columns give the spatial data, the coordinates of the centre ($X$, $Y$) and
the size of the area occupied by the members. In this column, we 
provide the semiaxes for OCC~1 as described above. For OCC~2 the two
numbers stand for the radii of the outer and the inner borders of the area of
density excess. For the Perseus-Auriga and Hyades groups, the sizes are the radii of areas
containing the selected candidates. $\olin{T}_l$ and $\olin{T}_b$ are the mean
tangential velocities corrected for differential Galactic rotation and Solar
motion with respect to the cluster centroid. For OCC~1, OCC~2, and
the Perseus-Auriga group they were used 
for the kinematic selection of members (see \S~\ref{memb_sec}). For the Hyades
group we computed $\olin{T}_l$ and $\olin{T}_b$ only with clusters close to
$l = 90^\circ$ ($l=44-124^\circ$) where $\olin{T_l}\approx-\olin{U}$.
The columns $\olin{U}$, $\olin{V}$, and $\olin{W}$ give the components
of space velocity corrected for differential Galactic rotation and Solar
motion with respect to the cluster centroid, and $n$ is the number of clusters
used for computing these velocity components. 
This number indicates, that only
for OCC~1 the space velocity was computed with all identified complex members.
In OCC~2, less than 30\% of its members can contribute to the determination
of the space velocity.  For the other groups, these values are merely illustrative.
Fig.\ref{occ_all_fig} shows relative location of the discussed cluster groups in
the $XY$ plane.

\begin{figure} 
\resizebox{\hsize}{!}
{\includegraphics[bb=56 198 538 638,clip=]{3764f10.ps.gz}}
\caption{The relative position of identified open cluster complexes and groups
in the $XY$ plane on a background of interstellar clouds revealed from data on our
sample of clusters (Fig.~\ref{av_fig}). Large symbols mark candidate
members for the Gould's Belt complex (triangles), OCC~2 (circles),
the Perseus-Auriga group
(diamonds), and the Hyades group (crosses). Small open circles  mark field
clusters. Large circles and the ellipse are apparent complex boundaries. 
}\label{occ_all_fig}
\end{figure}

Due to a relative proximity to the Sun and due to the completeness of kinematic data, OCC~1 
is the most prominent structure found in our sample. One of the important features of
this complex is its clustering in 3D-space. OCC~1 represents a density enhancement
which is about a factor three higher than the surrounding density of field T1 clusters 
(75 vs. 25 clusters per square kpc). It is clearly outlined in space and shows a
shape which is usually associated with  Gould's Belt. It is worth to note that
up to now Gould's Belt is studied only in connection with field OB stars
(Frogel \& Stothers~\cite{fros}, Torra et al.~\cite{tfg00}), local gas (see 
P\"oppel~\cite{pop97} for a detailed review), and local associations (Stothers \&
Frogel~\cite{stof}, de Zeeuw~\cite{dez}, and Bobylev~\cite{bobyl}), and there
was no attempt to study it in relation with young open clusters in a
systematic way.

\begin{table*}
\tabcolsep 3pt
\caption[]{Parameters of newly identified cluster complexes and groups}
\label{grpar_tab}
\begin{tabular}{lrcrrcrrrrrr}
\hline
\rule{0mm}{4mm}Complex,&$N$&$\log t$&\mc{1}{c}{$X_c$}&\mc{1}{c}{$Y_c$}&size&\mc{1}{c}{$\olin{T}_l$}&\mc{1}{c}{$\olin{T}_b$}&\mc{1}{c}{$\olin{U}$}&\mc{1}{c}{$\olin{V}$}&\mc{1}{c}{$\olin{W}$}&$n$\\
group   &          &   yr      &\mc{1}{c}{kpc} &\mc{1}{c}{kpc}&\mc{1}{c}{kpc}&\mc{1}{c}{km/s}&\mc{1}{c}{km/s}&\mc{1}{c}{km/s}&\mc{1}{c}{km/s}&\mc{1}{c}{km/s}         &    \\
\hline
OCC~1         &23    &$6.67-7.88$  & $-0.078$&$-0.053$ & $0.385,0.475$ &$-3.6\pm 1.4$ &$+1.0\pm 1.0$ &~$-4.0\pm 1.2$ &~$-0.2\pm 1.1$ &$+0.3\pm 0.6$&23\\   
OCC~2         &27    &$8.48; 0.09$& $0.000$&$0.000$   & $0.325,0.65$ &$+4.0\pm 0.8$ &$+2.9\pm 0.7$ &~$+1.9\pm 4.2$ &~$+3.3\pm 1.2$ &$+2.5\pm 1.3$& 8\\   
Perseus-Auriga&8     &$8.47; 0.09$& $-0.450$&$0.300$  & $0.26$ &$-12.6\pm 0.8$&$ +5.9\pm 1.2$&~$ +4.6\pm13.6$&$+21.1\pm 9.3$&$ +4.3\pm 4.3$& 2\\   
Hyades$^{1)}$ &9&$8.83; 0.21$& $0.080$&$0.540$   & $0.66$ &$+34.9\pm 3.7$&$- 0.4\pm 4.5$&$-34.4\pm 5.3$ &~$-0.8\pm 3.8$&$ +2.0\pm 2.6$& 4\\   
\hline
\end{tabular}\\
\rule{0mm}{3mm}
$^1$~\parbox[t]{16cm}{\footnotesize The mean tangential velocities were computed with 6 clusters
(without the Hyades, Praesepe, and ASCC~99)}\\
\end{table*}

Based on the present data we can regard the Gould's Belt complex of open clusters
as a physical system containing more than 20 clusters, occupying a flat elongated area
of size of about 0.8$\times$1$\times$0.2 kpc. The cluster ages range from a few to about 80
Myr. About half of the clusters are concentrated in a prominent group residing
in the region of the Orion association. They have the youngest ages (less the 25
Myr) among the OCC~1 clusters. The other clusters are distributed along the complex more
uniformly, and as a rule, they are older than 25 Myr. From their location, we find them
at the periphery of local molecular clouds with embedded associations which
expand from Vela to Perseus via Scorpio-Centaurus. Only at the top of the 
spatial distribution above the Galactic plane (in the region of the IC~4665 and 
Stephenson~1 clusters), OCC~1  emerges out of the molecular clouds. The complex
shows kinematics typical for young clusters, though it is more compact in the VPD
than field clusters. 

The first hint on possible cluster complexes was made by Lynga~(\cite{lyn82}), who 
identified three areas of high concentration of young ($\log t<7.5$) clusters in the
Perseus, Carina, and Sagittarius regions. Based on  comparative statistics of
populations of the suggested complexes and of the field, Janes et al.~(\cite{janea88}) 
found that the three complexes contain mainly clusters younger than 40 Myr, and
cover areas of order of 1.2...1.6 kpc. An extended search for clustering 
was undertaken by Efremov \& Sitnik~(\cite{efsi88}). Among
stellar associations and young clusters (earlier than b2 in the scale of 
Becker \& Fenkart~\cite{baf71}), they identify 17 groups of various sizes
(150...700 pc) containing at least 3 objects. They suggested that complexes
of young clusters both outline the spiral arms and fill inter-arm space 
in the Solar neighbourhood (see Efremov~\cite{efr95} for more details).

The presence of two cluster complexes and two more cluster groups in the Solar
neighbourhood, with only one being regarded as a young object implies that
such aggregates seem to be quite usual in the galactic disk. Therefore, it is not 
necessary to relate them to spiral arms or current star formation. If we do not
suggest an exclusive role for the Solar neighbourhood, we may expect to find footprints
of cluster groups (or their remnants) of different age everywhere in the
Galactic disk. With time, such groups should loose their members due to cluster 
evaporation,  become less prominent and will be detectable in the velocity space only.

We would like to stress that revealing the complexes and groups of clusters was not
an output of a systematic search for such objects in the Galactic disk but
rather the result of an attempt to understand the nature of irregularities
either in density  or in kinematic distribution which we found within the completeness
area of the cluster sample. Therefore, one may expect that a systematic search
in larger samples (which  perhaps are less complete but cover larger areas) will result in
identification of additional cluster agglomerations.

\section{Formation rate and lifetime of clusters}\label{ltime_sec}

One of the first extensive studies into the distribution of ages of open 
clusters and the estimation of their lifetime was carried out by Wielen~(\cite{w71}).
The statistics of cluster ages was based on data from the catalogues of
Lindoff~(\cite{lindoff}), and Becker \& Fenkart~(\cite{baf71}). 
Later studies  on this topic by Janes \& Adler~(\cite{janad82}),
Lyng{\aa}~(\cite{lyn82}), Pandey \& Mahra~(\cite{panma86}), Pandey et
al.~(\cite{panea87}), Janes et al.~(\cite{janea88}), Bhatt et al.~(\cite{bhapm})
and Battinelli \& Capuzzo-Dolcetta~(\cite{batin91}) used the approach proposed
by Wielen~(\cite{w71}), but considered  the larger list of clusters by
Janes \& Adler~(\cite{janad82}) as well as the third, forth, and fifth versions of the
Lund catalogue (Lyng{\aa}~\cite{lyn83}, \cite{lyn85}, \cite{lyn87}).  In comparison to
these studies, our sample contains at least twice as many clusters
with ages and distances determined by a uniform technique, and it includes many newly detected clusters.
Therefore, a revision of the previous findings and conclusions is reasonable and relevant.

Let $dN(t)=\psi(t)\,dt$ be the number of clusters formed in some interval of time
$(t,t+dt)$, where $\psi(t)$ is the cluster formation rate (CFR). Further, we assume
that the clusters decay with a probability $p=1/\tau$
where $\tau$ is the typical lifetime of these clusters. If $N(a)$ is the number
of clusters with age $a$, which are observed at the present time $t_1=t+a$,
we can relate $N(a)$ to the CFR via the equation
\begin{equation}
N(a) = N(t_1-a) \,e^{-a/\tau} = \psi(t_1-a)\,e^{-a/\tau}. \label{decay_eq}
\end{equation}
If clusters would never dissolve (i.e., $\tau = \infty$), then their age distribution
would reflect the temporal behaviour of the cluster formation rate. In contrast, for
a constant CFR the age distribution $N(a)$ is controlled by the decay process.
In the general case, the observed distribution of cluster ages  depends on both
factors, but in principle, it is possible to fix the processes. Since the
observed span of cluster ages (a few Gyr) is much shorter than the age of the Galactic
disk (over 10 Gyr) and since no indications of strong temporal variations are observed
for the local rate of star formation (see e.g. Scalo~\cite{scalo86}), it is reasonable to
assume that the formation rate of clusters is time-independent.
Analysing the age distribution of open clusters under the assumption of a constant CFR during the 
disk lifetime, Wielen~(\cite{w71}) found a cluster half-lifetime $t_{1/2}$ of 160 Myr, and
a cluster formation rate of $\psi_0=0.1$ kpc$^{-2}$Myr$^{-1}$. Here $t_{1/2}$ is the decay time
of one half of the clusters once formed, and it can be derived from eq.~(\ref{decay_eq}) as 
$t_{1/2}=- \tau\,\ln 1/2 \approx 0.7\,\tau$.
Using larger samples of clusters, the authors above obtained about the same
or somewhat smaller lifetimes of open clusters. From a comparison of the age parameters for
clusters of different richness classes and of different spatial location in the Galaxy,
a reasonable conclusion had been achieved that the cluster lifetime increases with the richness class
as well as outwards from the Galactic centre.

\begin{figure}[tb]
\resizebox{\hsize}{!}
{\includegraphics[bb=60 200 535 670,clip=]{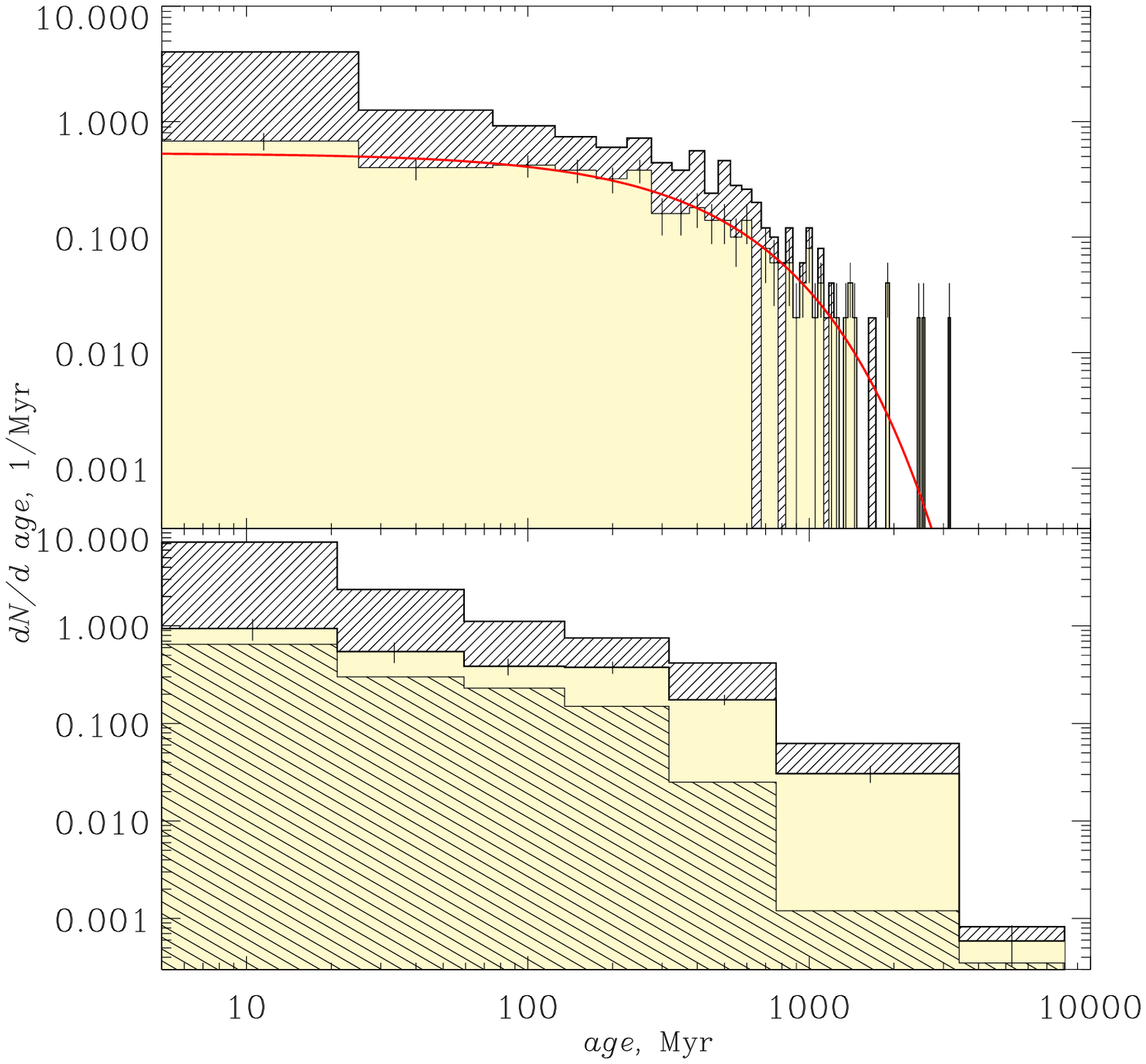}}
\caption{Distribution of open clusters versus age. For an easier comparison,
the distributions for the different samples are not normalised to unit area.
\textbf{Upper panel:} our total sample is shown as the hatched histogram.
The sample of field clusters within the completeness area is marked as the filled histogram,
and the solid curve marks the fitted age distribution. The age step is 50 Myr.
\textbf{Lower panel:} the same distributions as in the upper panel together 
with the age sample used by Wielen~(\cite{w71}).
The data from Wielen~(\cite{w71}) are shown as the backhatched histogram in
the foreground. The vertical bars correspond to Poissonian errors derived from
cluster counts. The binning is chosen same as in Wielen~(\cite{w71}).
}\label{dndt_fig}
\end{figure} 

Taking into account the statistical properties of our cluster sample, we computed
the CFR and the cluster lifetime for three cases. At first, we consider
the total sample of 652 clusters. In the second case, we include only those 259 
clusters which are located within the completeness area ($d_{xy}\le 0.85$ kpc). Finally, in order to 
eliminate possible biases in the age distribution due to the cluster complexes, we 
excluded their $1\sigma$-members from the second sample. This sample of so-called
field clusters contains 195 clusters.

In the upper panel of Fig.~\ref{dndt_fig} the distributions of clusters versus 
age are shown for the total sample as well as for the field clusters
within the completeness area. As expected, 
the total sample is biased towards young clusters. Compared to 
the distribution of field clusters, this results in a steeper average slope of the relation 
and, consequently, in smaller lifetimes. The curve obtained from fitting eq.~(\ref{decay_eq})
coincides well with the observed distribution of field clusters. The parameters of the fit
are given for all three cases in the first three rows of Table~\ref{cfr_tab}.
We conclude that the completeness issue has a strong impact onto the determination of 
CFRs and cluster lifetimes.  The exclusion of 
the complex members does not change the lifetime $\tau$ significantly, whereas 
the impact on $\psi$ is somewhat stronger: it decreases the CFR by about 20\%, but 
the difference is still within the mean errors.

\begin{table}
\centering
\caption[]{Cluster formation rates ($CFR$) and cluster lifetimes ($\tau$)}
\label{cfr_tab}
\begin{tabular}{rcccll}
\hline
No &$CFR$                   &$\tau$     &$N_{cl}$&Sample&Note\\
   &kpc$^{-2}$Myr$^{-1}$    &Myr        &        &      &    \\
\hline
 1& 0.10$\pm$0.01           &256$\pm$12 &652     & tot  &  \\
 2& 0.31$\pm$0.03           &327$\pm$25 &259     & cmp  &  \\
 3& 0.23$\pm$0.03           &322$\pm$31 &195     & fld  &  \\
\hline
 4& 0.10                    & 231       & 70     & w71  & 1\\
 5& 0.18                    & 144       & 112    & p86  & 2 \\
 6& 0.25                    & 100       & 213    & j88  & 3\\
 7& 0.45                    & 14        & 100    & b91  & 4\\
\hline
\end{tabular}\\
\rule{0cm}{0.2cm}
\parbox{\hsize}{\textbf{Sample:} No 1...3 are the present study, No 4...7 are
taken from the literature;
tot -- the total sample; cmp -- the sample of clusters within the completeness area; 
fld -- the sample of field clusters within the completeness area;
w71 -- Wielen~(\cite{w71}); p86 -- Pandey \& Mahra~(\cite{panma86}); j88 --
Janes et al.~(\cite{janea88}); b91 -- Battinelli \&
Capuzzo-Dolcetta~(\cite{batin91}).
}\\
\parbox{\hsize}{\textbf{Notes:} 1 -- the Becker \& Fenkart~(\cite{baf71}) sample;
2 -- the Lund catalogue, 3rd edition; 3 -- the Lund catalogue, 5th edition;
4 -- clusters with $d\le 2$kpc \& $V_{integr} \le -4.5$ mag.}
\end{table}

In the bottom panel of Fig.~\ref{dndt_fig} we show the same distributions  
together with results from Wielen~(\cite{w71}) based on the Becker \& Fenkart~(\cite{baf71}) sample.
In its shape, the distribution used by Wielen~(\cite{w71}) is in  better agreement with
the data of the total sample, but even so, there is a pronounced deficiency of older clusters
which play an important role in the determination of cluster lifetime. This is the reason
for the smaller lifetime derived by Wielen~(\cite{w71}), given in Table~\ref{cfr_tab}, too.

Table~\ref{cfr_tab} also includes the corresponding findings by other authors 
from the 80s and 90s. 
To ease comparison, we transformed $t_{1/2}$ to $\tau$, if the cluster
lifetimes were not explicitly given in the original publication.
Not only do the determinations of $\tau$ differ strongly among themselves,
they are far off Wielen's results, and even farther from ours.
We explain the disagreement of our results with the results of 
Pandey\& Mahra~(\cite{panma86}),
and Janes et al.~(\cite{janea88}), based on the Lund Catalogue, by the considerable increase of data on 
clusters older than 250 Myr in the last decade. For example, the number of clusters with ages
of $250\ldots3000$ Myr within 0.85 kpc from the Sun has been increased by a factor of 6
due to the newly discovered clusters. In comparison, the total number of clusters
has grown from 1150 in the 5th edition of the Lund Catalogue to about 1700 clusters 
in Dias et al.~(\cite{dlam}) i.e., by only a factor of 1.5.
Concerning the extremely low lifetime $\tau$ derived by 
Battinelli \& Capuzzo- Dolcetta~(\cite{batin91}, the selection effect is even stronger.
Since they considered only the brightest clusters, their sample is overabundant 
on young objects, and the lifetime is strongly underestimated.

The footprints of selection effects become also visible in the results for the CFR
in Table~\ref{cfr_tab}. Unlike the lifetime determination which mainly depends on 
a realistic relation between numbers of young and old cluster,
the CFR results are more influenced 
by overall incompleteness of the sample studied. The increase of the CFRs computed with our samples
correlates clearly with an increasing completeness of the data (cf. Table~\ref{cfr_tab}).
A similar correlation can be observed for the CFR results derived by 
Wielen~(\cite{w71}), Pandey \& Mahra~(\cite{panma86}), and Janes et al.~(\cite{janea88}).
The good agreement between our CFR and that from Janes et al.~(\cite{janea88}) is
expected, since we use the same basic source, the 5th edition of the Lund Catalogue.
On the other hand, we believe that the considerable disagreement in the CFR derived
by Battinelli \& Capuzzo-Dolcetta~(\cite{batin91}) and all others is a 
consequence of the underestimation of ages of their young clusters. It is well known
that the turn-off age calibrations are biased in the case of young clusters (a few tens of Myr or younger)
due to the rather steep Main Sequences of early-type stars towards younger ages.
Neglecting this bias, one underestimates the age spread within a sample of young clusters,
and hence, overestimate their abundance.

Taking the cluster lifetime derived above and an age $t_1$ for the Galactic disk of about 
10-12 Gyr,  we compute the total number of cluster generations  $n_g=t_1/\tau$ as being of the
order of 30-40. Provided that a typical open cluster consists of $n^*_o=1000$
stars at its birth and with the surface density of clusters being $\Sigma=114$ kpc$^{-2}$
(see \S~\ref{z_sec}), the total surface density $n^*$ of disk stars and their remnants
which have passed an open cluster member phase is $n_g\,n^*_o\,\Sigma\approx4\times10^6$ kpc$^{-2}$.
This number is in a good agreement with another estimate coming from the 
derived CFR i.e., $n^*=n^*_o\,\psi\,t_1\approx3.6\times10^6$.
Compared with the local density of disk stars of about $7\times10^7$ kpc$^{-2}$,
we find that less than 6\%
of the total stellar population of the Galactic disk
is genetically connected with classical open clusters. On the other hand, this estimate is
based on assumptions and extrapolations which at present can not be checked in detail. Therefore, 
we can only conclude that that portion of the disk stellar population, which had been 
originated in open clusters, is relatively small, of the order of 10\%. 

This estimate is implicitly supported by another statistics available from the
ASCC-2.5. Let us count all stars which are brighter than $V=9$ mag i.e., a
magnitude range where  99\% of stars have spectral classification and, practically, where all clusters are
identified in the ASCC-2.5. Among the intrinsically bright population
(supergiants, O -- B3 stars), we then computed the ratio of $1\sigma$ and
$2\sigma$ cluster members to field stars. It turns out that  about 10\% 
of these 3770 extremely luminous stars are members  of open clusters. Our
results support
the ``pessimistic'' estimate of 10\% obtained by Miller \&
Scalo~(\cite{misc}) from a much poorer statistics.

The rather low input of open clusters to the stellar population of the Galactic disk
can be explained by their high infant mortality, reducing  generations of
newly born clusters by a half (Lamers et al.~\cite{lamea}). Even so, this is not
sufficient to account for all Population~I stars. Thus, the current statistics
support the early conclusion by Miller \& Scalo~(\cite{misc}) that about
65\% of the disk stars should have their origins in the association reservoir.

\section{Conclusions}\label{concl_sec}

In this paper we studied unbiased properties of the local cluster
population taking advantage of a homogeneous set of open cluster parameters
derived from the all-sky census of cluster stars up to $V$ = 12.5. 
The revised cluster membership, the implementation of accurate kinematic and 
photometric data supplemented by radial velocities and spectral classification
available at the moment, allowed us to extend the completeness area of open cluster data
in the optical up to a distance of 0.85 kpc. The resulting sample was the basis for
the determination of spatial, kinematical and evolutionary parameters of
cluster population in the local Galactic disc.

Although the spatial distribution of clusters on scales comparable to inter-arm distances
($d_{xy}\approx 2$ kpc) is compatible with a model of a constant surface density, the
distribution is not uniform. There are two main factors influencing the observed
distribution in the optical. At first, the cluster can form groups which are
revealed in the age distribution as well as in the space and velocity
distributions. We found three cluster complexes in the Solar neighbourhood
affecting significantly the apparent density pattern. The other effect
influencing the apparent distribution of nearby open clusters in the optical is
a patchy interstellar extinction due to local screening clouds. Also, these have
a considerable impact on the observed  distribution of clusters at large
distances which we can see only through transparency windows. It is not excluded
that some features of the local disk (like the Local arm) which have  been discussed
for a long time, are only virtual details which appear due to an uneven
distribution of the reddening.  The vertical distribution of the cluster sample
turned out to be extremely flat, with a scale height of about 50\ldots60 pc, and
it does not depend on age at least for cluster ages up to 1~Gyr. We found the
symmetry plane of open clusters to be $Z_0=-22\pm4$ pc, and the total density of
clusters in the symmetry plane is $D(Z_0)=1015$ kpc$^{-3}$. The spatial
distribution of younger clusters is compatible with a flat grand design of a
spiral pattern with a pitch angle of $p=-5^\circ\ldots-6^\circ$.

The kinematical parameters of the cluster sample are comparable to those 
of young field stars in the Solar neighbourhood though, the clusters cover distances 
which are typically larger by a factor of two than Hipparcos-based samples of 
field stars. This confirms that, in general, the observed cluster 
population is young. A combination of accurate cluster ages and kinematical parameters
provides a possibility to study the temporal variation of the cosmic velocity dispersion.
We found that in average, the dispersion of each velocity component increases 
by a factor of two during a period of 3 Gyr. The local velocity field within about 1.5 kpc
can be sufficiently well described by the linear model of differential rotation.
But similar to the density distribution, the regular velocity pattern can be broken by
groups of clusters with significantly different motions. So, we found a group of nine 
older clusters showing a large velocity component in the Galactic anti-centre direction.

Fluctuations in the spatial and velocity distributions can be attributed
to the existence of three open cluster complexes containing up to a few tens of clusters
of similar ages. The youngest complex, OCC~1 ($\log t<7.9$), is a structure of about
1~kpc in size inclined to the Galactic plane by 19$^\circ$. Highly probable, it is a 
signature of Gould's Belt which has generally been considered in the literature
to be an aggregation of the
youngest field stars, gas and associations. The most populated complex, OCC~2, includes
clusters of moderate age ($\log t\approx8.45$). The clusters of the Perseus-Auriga
group have the
same age but show different and compact kinematics. They are also very compact
in their location and seen in voids of the Perseus-Auriga clouds. Possibly, we
see only a part of a larger structure covered by clouds. The oldest ($\log
t\approx8.85$) and the most loose group was detected due to a large peculiar
motion (i.e., velocities corrected for Galactic rotation and Solar motion) of
about 25-45 km/s in the Galactic anticentre direction.  We note that the Hyades
and Praesepe have about the same age ($\log t$ = 8.9) and they have an $U$
velocity component of -34.1 and -30.9 km/s, respectively. Nevertheless, the real
existence of the Hyades moving group of open clusters cannot be proven due to
the lack of radial velocities of the suggested candidates.

From the parameters of spatial distribution, we estimated a total number of 10$^5$
open clusters currently in the Galactic disk. Cluster lifetime and
formation rate obtained from the age distribution of field clusters within the completeness area
are found to be 322$\pm$31 Myr and $0.23\pm0.03$ kpc$^{-2}$Myr$^{-1}$, respectively.
Assuming a typical open cluster of the Pleiades type, one derives the total surface 
density of disk stars passed through the phase of open cluster members to be
about $4\times10^6$ kpc$^{-2}$. Compared to the local density of disk stars of
about $7\times10^7$ kpc$^{-2}$, one obtains that the input of open clusters
into the total population of the Galactic disk is about 6\%.

\begin{acknowledgements} 
This work was supported by DFG grant 436~RUS~113/757/0-1, RFBR grant
03-02-04028. We acknowledge the use of the Simbad database  and the VizieR
Catalogue Service operated at the  CDS, and the WEBDA facility at Observatoire de
Geneva. We are indebted to A.V. Loktin for making his data available prior
to publication, and V.S. Avedisova for useful discussions. We are
grateful to the anonymous referee for his/her useful comments.
\end{acknowledgements}

\end{document}